\documentclass[12pt,a4paper]{article}
\usepackage{graphicx}
\usepackage[T1]{fontenc}
\usepackage[utf8]{inputenc}
\usepackage{textcomp}
\usepackage[sc,osf]{mathpazo}
\usepackage{a4wide}  
\usepackage{latexsym,amsthm,amsfonts,amsmath,mathrsfs,amssymb}
\usepackage{dsfont}
\usepackage{xcolor}
\usepackage{accents}
\usepackage{physics}   
\usepackage{amsmath}  
\usepackage{derivative}
\usepackage{comment}
\usepackage{cancel}  

\usepackage[nosort]{cite}
\usepackage{booktabs} 
\usepackage[unicode,implicit]{hyperref}
\hypersetup{%
  pdftitle = {Derivation of the Smarr formula from the Komar charge in
    Einstein-nonlinear electrodynamics theories and
    applications to regular black holes}
  pdfkeywords = {gravity,  symmetries, conserved charges, Komar charge,
    Komar intergral,  non-linear electrodynamics, Smarr formula, black holes},
  pdfauthor   = {Gabriele Barbagallo and Tom\'as Ort\'{\i}n},
  plainpages  = true,
  colorlinks  = true,
  citecolor   = blue,
  urlcolor    = red,
  linkcolor   = black
}
\newcommand{\hepth}[1]{{\tt
\href{http://www.arXiv.org/abs/hep-th/#1}{hep-th/#1}}}
\newcommand{\grqc}[1]{{\tt
\href{http://www.arXiv.org/abs/gr-qc/#1}{gr-qc/#1}}}

\newcommand{\arxiv}[1]{{\tt arXiv:\href{http://www.arXiv.org/abs/#1}{#1}}}
\newcommand{\doi}[1]{{\tt DOI:\href{http://dx.doi.org/#1}{#1}}}

\allowdisplaybreaks

\makeatletter
\@addtoreset{equation}{section}
\makeatother

\begin{document}

\begin{flushright}
\small
IFT-UAM/CSIC-26-43\\
\texttt{arXiv:2605.02813 [hep-th]}\\
May 4\textsuperscript{th} 2026\\
\normalsize
\end{flushright}

\vspace{.2cm}

\begin{center}

  {\Large {\bf Derivation of the Smarr formula from the Komar charge
      \\[.5cm] in Einstein-nonlinear electrodynamics theories \\[.5cm]
      and applications to regular black holes }}
 
\vspace{1cm}

\renewcommand{\thefootnote}{\alph{footnote}}

{\sl Gabriele Barbagallo}\footnote{Email: {\tt gabriele.barbagallo[at]estudiante.uam.es}}
{\sl and Tom\'{a}s Ort\'{\i}n}\footnote{Email: {\tt Tomas.Ortin[at]csic.es}}

\setcounter{footnote}{0}
\renewcommand{\thefootnote}{\arabic{footnote}}
\vspace{1cm}

{\it\small Instituto de F\'{\i}sica Te\'orica UAM/CSIC\\
C/ Nicol\'as Cabrera, 13--15,  C.U.~Cantoblanco, E-28049 Madrid, Spain}

\vspace{1cm}

{\bf Abstract}
\end{center}

\begin{quotation} {\small We construct the generalized Komar charge of
    generic, non-linear theories of electrodynamics (NLED) in 4
    dimensions coupled to Einstein gravity.  The contribution of the
    dimensionful coupling constant present in all these theories is
    obtained by promoting it to a dynamical field which is forced to
    be constant on-shell by a Lagrange multiplier.  We use this charge
    to derive a Smarr formula for asymptotically-flat black-hole and
    soliton solutions of these theories that includes the contribution
    of the coupling constant.  Previously, this contribution had been
    found using homogeneity arguments.  We test our results on a broad
    class of Einstein--NLED theories and analyze in detail the
    thermodynamics of the regular Bardeen black hole using the
    conservation of the generalized Komar charge to understand the
    regularity of regular black holes inside the event horizon.  }
\end{quotation}

\newpage
\pagestyle{plain}

\tableofcontents

\newpage

\section{Introduction}

Models of nonlinear electrodynamics (NLED) have been extensively
investigated as effective descriptions of strong-field or
quantum-corrected electromagnetic phenomena and as probes of physics
beyond Maxwell's theory. These models have also found important
applications in gravity, cosmology and string theory; for example,
Born--Infeld theory emerges in string theory in the low-energy
effective description of open strings and D-brane dynamics.

NLED theories coupled to pure Einstein gravity (Einstein--NLED
theories) have attracted attention recently because, some of them,
admit regular, charged, soliton and black-hole solutions \cite{Dymnikova:1992ux, Ayon-Beato:1998hmi, Ayon-Beato:1999kuh, Ayon-Beato:1999qin, Ayon-Beato:2000mjt,Ayon-Beato:2004ywd, Bronnikov:2000vy, Dymnikova:2015hka,Bronnikov:2022ofk, Bronnikov:2017sgg, Dymnikova:2004zc}.
Our goal in this work is to get a better understanding of the
thermodynamic properties of all the black-hole solutions of these
theories (in particular, of the Smarr formula and the first law) that
may explain how and when they can be fully regular.

In generic Einstein--NLED theories, the presence of a dimensionful coupling
constant (that we will always call $\alpha$) modifies the Smarr formula
\cite{Smarr:1972kt}. The zeroth and first laws of black hole thermodynamics
for this class of theories were established in the seminal work
Ref.~\cite{Rasheed:1997ns}, which also provided one of the earliest attempts
to generalize the Smarr relation. That attempt, however, was ultimately
unsuccessful, since the standard Smarr formula was found not to hold in its
usual form. In later studies
\cite{Breton:2004qa,Gonzalez:2009nn,Yi-Huan:2010jnv,Gunasekaran:2012dq,Diaz-Alonso:2012lkh,Zhang:2016ilt,Fan:2016hvf,Hu:2018njr},
homogeneity arguments similar to those used originally in
Ref.~\cite{Smarr:1972kt} were employed to extend the Smarr relation by
introducing an additional term associated with the coupling constant, which is
treated as a thermodynamic variable, and its new conjugate quantity. This
strategy parallels the extended phase-space treatment in which the
cosmological constant is interpreted as a thermodynamic pressure
Ref.~\cite{Kastor:2009wy} (see also the review Ref.~\cite{Kubiznak:2016qmn}
and references therein).  Remarkably, in Ref.~\cite{Gulin:2017ycu}, using the
techniques explained in Ref.~\cite{Townsend:1997ku}, the authors derived the
generalized Smarr formula for Einstein--NLED theories in $d=4$ dimensions, in
which the extra term related to the coupling constant turns out to be
proportional to the integral of the trace of the energy-momentum tensor (see
also Ref.~\cite{Bokulic:2021dtz,Balart:2017dzt}).

As first shown in Ref.~\cite{Bardeen:1973gs}, the Komar charge
\cite{Komar:1958wp} can be also used to obtain the Smarr formula, which
relates quantities defined at infinity with quantities defined on the event
horizon, in a more geometrical way. This derivation uses the fact that it is
closed on-shell and its integral over topologically equivalent
surfaces\footnote{That is: surfaces that can be smoothly deformed into each
  other without crossing any singularity in the relevant background. Since the
  derivation is actually based on the Stokes theorem, it is enough that the
  surfaces are cobordant with no singularities in the volume their union is
  the boundary of. Thus, this technique can be used with horizons with
  topology different from the sphere at infinity, such as the horizons of
  black rings considered in Ref.~\cite{Barbagallo:2025qdy}.} always gives the
same value (\textit{i.e.}~it satisfies a Gauss law), and the fact that the
sphere at infinity is topologically equivalent to the event horizon.

In order to generalize this technique to theories other than pure General
Relativity it is necessary, first, to derive a generalization of the Komar
charge which is closed on-shell in the theory under consideration. For the
(cosmological) Einstein--Maxwell theory a \textit{generalized Komar charge}
was proposed in Refs.~\cite{Carter:1973rla,Magnon:1985sc,Bazanski:1990qd}, but
only for the asymptotically flat case a Smarr formula was derived in the first
of these references and the cosmological case was not considered until more
recently in Refs.~\cite{Kastor:2008xb,Kastor:2010gq}. In
Ref.~\cite{Ortin:2021ade} it was proposed that other dimensionful constants
appearing in the action may also be treated as thermodynamical variables and
in Ref.~\cite{Meessen:2022hcg} it was shown following a proposal made in
Ref.~\cite{Liberati:2015xcp}, how to construct in a systematic way the
generalized Komar charges of theories containing such constants in the context
of Wald's formalism \cite{Lee:1990nz,Wald:1993nt}, promoting them to scalar
fields and adding Lagrange multiplier terms to the action that ensure they are
constant on-shell. The same techniques, supplemented with those proposed in
Refs.~\cite{Elgood:2020svt,Elgood:2020mdx,Elgood:2020nls,Mitsios:2021zrn,Ortin:2022uxa}
can be used to properly include the magnetic charges and to derive the first
law.

Einstein--NLED theories typically contain a dimensionful constant that we will
call $\alpha$ that can be treated in this way and our goal in this work is to
use the techniques mentioned above to derive the Smarr formula and the first
law for these theories always in $d=4$ dimensions. Since one can also
integrate the generalized Komar charge inside the event horizon we will use it
to explore the mechanism that maintains the value of the integral constant
when the surface of integration shrinks to a point in absence of a
singularity.

In order to illustrate this last point, let us consider an example in
4-dimensional Maxwell electrodynamics. Maxwell's equations can be written in
the form\footnote{The symbol $\doteq$ is used to denote identities that
	may only hold on-shell.}

\begin{equation}
  \label{eq:Maxwellequations}
d\star F \doteq\mathbf{J}\,,  
\end{equation}

\noindent
where the electric current is represented here by its dual: the 3-form current
$\mathbf{J}$. The integrability condition of the above equation is the
conservation of the current, which in this language is

\begin{equation}
  \label{eq:dJ=0}
  d\mathbf{J}
  \doteq
  0\,.
\end{equation}

\noindent
The electric charge contained in the 3-dimensional spatial volume $\Sigma^{3}$
can be defined as

\begin{equation}
q = \tfrac{1}{4\pi} \int_{\Sigma^{3}}\mathbf{J}\,.  
\end{equation}

\noindent
Maxwell's equations Eq.~(\ref{eq:Maxwellequations}) and Stokes' theorem imply
that the charge contained in $\Sigma^{3}$ is also the flux of the electric
field across the boundary of the volume $\partial \Sigma^{3}$ (\textit{Gauss
  law})

\begin{equation}
  q
  =
  \tfrac{1}{4\pi}\int_{\partial \Sigma^{3}}\star F\,,
\end{equation}

\noindent
which is an expression that we can take as a convenient alternative definition
of electric charge when we do not know the source $\mathbf{J}$.

If the charge distribution is confined to a bounded region, the 2-form charge 

\begin{equation}
  \mathbf{Q}\equiv \tfrac{1}{4\pi}\star F\,,
\end{equation}

\noindent
is closed on-shell in the exterior of the distribution 

\begin{equation}
  \label{eq:dQ=0}
d\mathbf{Q} \doteq 0\,,  
\end{equation}

\noindent
and the charge $q$ can be computed over any closed 2-dimensional
surface containing the charge distribution. For the sake of
simplicity, let us consider the spatial volume $\Sigma^{3}_{r}$ contained
between the 2-sphere at spatial infinity, $S^{2}_{\infty}$ and a
2-sphere of radius $r$ enclosing the charge distribution, $S^{2}_{r}$ so that

\begin{equation}
  \label{eq:Sigma3r}
\partial \Sigma^{3}_{r} = S^{2}_{\infty} \cup S^{2}_{r}\,.  
\end{equation}

\noindent
Integrating Eq.~(\ref{eq:dQ=0}) over $\Sigma^{3}$ and using Stokes' theorem we
find that

\begin{equation}
q = \int_{S^{2}_{\infty}}\mathbf{Q}  \doteq \int_{S^{2}_{r}}\mathbf{Q}\,,
\end{equation}

\noindent
for any $r$. Actually, we would obtain the same result for any 2-dimensional
surface topologically equivalent to $S^{2}_{\infty}$ or $S^{2}_{r}$, that is:
surfaces that can be smoothly deformed into each other without crossing any
singularity (source) in the relevant background.

Now, if the sphere $S^{2}_{r}$ could be contracted to zero radius without
meeting any singularity, the value of the integral would be identically zero
and we would obtain a contradictory result. Thus, there must be a singularity
at the origin at which the vacuum Maxwell equations are not satisfied, which
we can identify as a point-like charge.

The existence of a singularity, however, relies on the assumption that
the spheres can be contracted to a point, but this may not be the case
in topologically non-trivial (wormhole-like) spacetimes in which the
Misner--Wheeler phenomenon of ``charge without charge''
\cite{Misner:1957mt} takes place.

We conclude that, in absence of non-trivial topology, there are no charged,
regular, solitons in the vacuum Maxwell theory.

How is this conclusion modified in the presence of a charge distribution
$\mathbf{J}\neq 0$? Since, by consistency, this current is closed on-shell,
Eq.~(\ref{eq:dJ=0}), locally, we should be able to write
$\mathbf{J}\doteq d \mathbf{S}$ for some 2-form $\mathbf{S}$ and construct a
new on-shell closed 2-form charge

\begin{equation}
  \mathbf{Q}' = \star F -\mathbf{S}\,,
  \hspace{1cm}
  d\mathbf{Q}' \doteq 0\,.
\end{equation}

This is the point of view adopted, for instance, in
Refs.~\cite{Ballesteros:2023muf,Ballesteros:2024prz} and, while correct, it
has an important shortcoming because, in general, $\mathbf{S}$ cannot be
written as a local function of the fields and derivatives\footnote{As shown in
  Ref.~\cite{Wald:1990mme}, this is possible when a quantity is identically
  (off-shell) closed.} which we later evaluate on a solution, but only as a
2-form on our manifold obtained only after $\mathbf{J}$ is evaluated on a
solution. Its form, is, therefore, absolutely solution-dependent and, in this
situation, it can be more advantageous to work directly with the original
2-form charge $\mathbf{Q}=\tfrac{1}{4\pi}\star F$ which is no longer closed
but satisfies

\begin{equation}
d\mathbf{Q} \doteq \tfrac{1}{4\pi} \mathbf{J}\,,  
\end{equation}

\noindent
as in Ref.~\cite{Ballesteros:2025wvs}.

Integrating over the hypersurface $\Sigma^{3}_{r}$ defined above with boundary
Eq.~(\ref{eq:Sigma3r}) an applying Stokes' theorem, we find that the total
charge of the spacetime, $q$ is given by

\begin{equation}
  \label{eq:QversusJ}
  q
  =
  \int_{S^{2}_{\infty}}\mathbf{Q}
  \doteq
  \int_{S^{2}_{r}}\mathbf{Q}
  +\tfrac{1}{4\pi}\int_{\Sigma^{3}_{r}} \mathbf{J}\,.
\end{equation}

Now, in the $r\to 0$ limit, if the total charge of the spacetime $q$ is given
by the volume integral, the surface integral over $S^{2}_{r}$ can consistently
vanish in this limit and neither singularities nor non-trivial topologies are
necessary.

The result that a total charge $q$ can be sourced by a completely regular
``charged matter'' distribution is hardly surprising, but the situation becomes
much more interesting when the 3-form current $\mathbf{J}$ is produced by the
Maxwell field itself. This will happen when the Maxwell action is modified by
the addition of non-linear terms, which, in $d=4$, generically requires the
introduction of the dimensionful constant $\alpha$ mentioned
above.\footnote{In $d=5$ dimensions, the Maxwell action can also be modified
  by a Chern--Simons 5-form $F\wedge F\wedge A$, as in minimal 5-dimensional
  supergravity \cite{Cremmer:1980gs}. The addition of this term does not
  require a new dimensionful coupling constant.} The result is a theory of
NLED in which the electric field corresponding to a total charge $q$ can be
regular without the help of ``charged matter'' or non-trivial topology.

An interesting example of this phenomenon is provided by the theory proposed
by Born and Infeld in Ref.~\cite{Born:1934gh}, based on the action (here in a
general metric background)

\begin{equation}
  \label{eq:BIaction}
  \begin{aligned}
    S_{BI}[A]
    & =
    \int d^{4}x \alpha\left\{
      \sqrt{|g|}
      -\sqrt{-\mathrm{det}\left(g_{\mu\nu}+\frac{1}{\sqrt{\alpha}}F_{\mu\nu}\right)}
      \right\}
    \\
    & \\
    & =
    \int d^{4}x\sqrt{|g|} \alpha\left\{
      1
      -\sqrt{1+\frac{1}{2\alpha}F^{2}
      -\frac{1}{16\alpha^{2}}\left(F\star F\right)^{2}}
      \right\}\,,      
  \end{aligned}
\end{equation}

\noindent
which approaches Maxwell's when $\alpha\to \infty$.

For fields such that $F\star F=0$, the equation of motion takes the simple form

\begin{equation}
  \label{eq:BIequations}
  d\left(  a \star F \right) \doteq 0\,,
\hspace{1cm}
 a=\frac{1}{\sqrt{1+\frac{1}{2\alpha}F^{2}}}\,,
\end{equation}

\noindent
and the electric charge can be defined as the integral of the on-shell closed
2-form charge

\begin{equation}
  \mathbf{Q}
  =
  \tfrac{1}{4\pi}a\star F\,.
\end{equation}

If we consider a spherically-symmetric field  with total electric charge $q$,
the Gauss law implies that

\begin{equation}
a\star F = q \omega_{(2)}\,, 
\end{equation}

\noindent
where $\omega_{(2)}$ is the volume form of the 2-sphere of unit radius. In
Minkowski spacetime in spherical coordinates,

\begin{equation}
  aF_{rt} = -\frac{q}{r^{2}}\,
  \,\,\,\,\,\,
  \Rightarrow
  \,\,\,\,\,\,
  \left\{
      \begin{array}{rcl}
        F_{rt} & = -{\displaystyle\frac{q}{\sqrt{r^{4}+\frac{q^{2}}{\alpha}}}}\,,\\
               & \\
        a & =\sqrt{1+\frac{q^{2}}{\alpha r^{4}} }\,, \\
      \end{array}
    \right.
\end{equation}

\noindent
The radial electric field $F_{tr}$ approaches the Coulomb field when
$\alpha\to \infty$ and when $r\to \infty$ but takes the finite value
$F_{rt}=-\sqrt{\alpha}$ when $r\to 0$. The screening factor $a$ goes to $1$ at
infinity but diverges at the origin and so does what we may call the radial
``electric displacement field'' $aF_{rt}$.

The Born-Infeld equations (\ref{eq:BIequations}) can also be written as the
Maxwell equations (\ref{eq:Maxwellequations}) with a 3-form current of the form

\begin{equation}
\mathbf{J} = -\frac{da}{a}\wedge \star F =  q da^{-1}\wedge \omega_{(2)}\,, 
\end{equation}

\noindent
whose integral over $\Sigma^{3}_{r}$ is

\begin{equation}
  \tfrac{1}{4\pi}\int_{\Sigma^{3}_{r}}\mathbf{J}
  =
  q\left[1 -a^{-1}(r)\right] \stackrel{r\to 0}{\longrightarrow} q\,,
\end{equation}

\noindent
which explains the absence of a singularity at the origin in the electric
field. Notice that this current can only be written a total derivative after
we evaluate it on a solution.


The same chain of arguments can be used in the context of theories of
gravity to investigate the existence of particle-like globally regular
solutions. The standard Komar charge 2-form associated to a
Killing vector $k=k^{\mu}\partial_{\mu}$

\begin{equation}
  \label{eq:K0def}
  \mathbf{K}_{0}[k]
  =
  -\frac{1}{16\pi G_{N}^{(4)}}\star d\hat{k}\,,
\end{equation}

\noindent
where $\hat{k}=k_{\mu}dx^{\mu}$, is closed on-shell in pure General
Relativity (GR). In stationary, asymptotically-flat spacetimes, if $k$
is the timelike Killing vector that generates time translations and
has unit norm at spatial infinity

\begin{equation}
  \int_{S^{2}_{\infty}} \mathbf{K}_{0}[k] = M/2\,,
\end{equation}

\noindent
where $M$ is the ADM mass. The fact that one obtains the same result
integrating over any other sphere of finite radius as long as we do
not cross any singularities when we deform one into the other is not
emphasized as often as it deserves to be.  In particular, it implies
that if the radius of the sphere can be contracted to zero we find the
same contradiction as in the electromagnetic case and we arrive at the
conclusion that there must be a singularity or non-trivial topology in
the 3-dimensional spacelike hypersurface connecting the 2-sphere at
infinity with the surface over which we are integrating. This is what
happens in typical black-hole spacetimes in which the hypersurface can
extend all the way to another asymptotic region or reach a
singularity. 

This proves that no massive topologically-non-trivial solitons exist in pure
General Relativity, a result which was originally obtained in far less
transparent ways,\footnote{See, for instance, the historical review and
  references cited in Ref.~\cite{Gibbons:2013tqa} and the discussion in the
  introduction of Ref.~\cite{Ballesteros:2024prz}.} but, since the argument is
independent of the existence of event horizons, it applies equally well to the
(non-) existence of regular black holes with a finite (topologically-trivial)
interior in General Relativity.\footnote{This argument does not preclude the
  existence of regular black holes with the topology of the one found in the
  context of string theory in Ref.~\cite{Cano:2018aod}, but the interior of
  the regular black holes in Einstein--NLED we are interested in is supposed
  to be topologically trivial with 2-spheres that can be contracted to a
  point.}

This result can be generalized to theories containing fields with standard
couplings \cite{Ballesteros:2024prz} in which the standard Komar charge is
generalized with the addition of other 2-forms that make it on-shell
closed. However, in some cases, the matter couplings allow for regular
solutions of trivial topology. In this context this could be explained by the
contribution of the new terms present in the generalized Komar charge. In
Ref.~\cite{Ballesteros:2024prz} it was shown that when the solutions satisfy a
\textit{generalized symmetric ansatz} (\textit{i.e.}~some of its fields are
invariant under a spacetime symmetry up to a global symmetry of the theory as
in Ref.~\cite{Yazadjiev:2024rql})\footnote{This is usually referred to as an
  example of the phenomenon of ``symmetry non-inheritance,''
  \cite{Smolic:2015txa} but can also be viewed as a more general realization
  of a symmetry.} there are additional contributions in the generalized Komar
charge that can solve the contradiction one finds when the integration surface
contracts to a point.

However, the point of view proposed in Ref.~\cite{Ballesteros:2025wvs}
according to which one should deal with a generalized Komar charge which is
not closed, can be more appropriate to describe the situation because the
additional terms in the generalized Komar charge can only be fully determined
after evaluation on solutions, as in the NLED model we have just discussed.

In this work we want to study from this point of view $d=4$ general
Einstein--NLED theories. As we are going to show, the on-shell closed
generalized Komar charges of these theories can be written in the form

\begin{equation}
\label{k0kned}
	\mathbf{K}[k]\equiv \mathbf{K}_{0}[k]+\mathbf{K}_{\rm NLED}[k]\,,
\end{equation}

\noindent
where $\mathbf{K}_{0}[k]$ is the standard Komar charge in Eq.~(\ref{eq:K0def})
and $\mathbf{K}_{\rm NLED}[k]$ contains the contributions the electromagnetic
and coupling constant terms. Some of the terms in $\mathbf{K}_{\rm NLED}[k]$
can only be computed after the evaluation of the 3-form they come from on a
solution and, therefore, according to the previous discussion and examples,
for some purposes it will be more convenient to use the equation

\begin{equation}
  \label{eq:nonconservationofK0}
  d\mathbf{K}_{0}[k]
  \doteq
  \mathbf{Z}[k]\,,  
\end{equation}

\noindent
where $d\mathbf{K}_{\rm NLED}[k] = -\mathbf{Z}[k]$ defines
$\mathbf{K}_{\rm NLED}[k]$ after $\mathbf{Z}[k]$ is evaluated on a
solution.\footnote{A similar relation that explains the existence of the
  massive, charged, globally regular, regular, horizonless solutions of
  $\mathcal{N}=1$, $d=5$ supergravity known as \textit{fuzzballs} was found by
  Gibbons and Warner in Ref.~\cite{Gibbons:2013tqa}. See also
  Ref.~\cite{Barbagallo:2025qdy}.}

We will use this generalized Komar charge to make provide a first-principles
derivation of the Smarr formula that does not rely on homogeneity arguments
and Eq.~(\ref{eq:nonconservationofK0}) to study the presence or absence of
singularities or non-trivial topology along the lines we have followed in the
examples provided in this introduction. Thus, we will integrate
Eq.~(\ref{eq:nonconservationofK0}) for the timelike Killing vector $k$,
deriving the equation

\begin{equation}
  M/2
  =
  \int_{S^{2}_{\infty}}\mathbf{K}_{0}[k]
  \doteq
  \int_{S^{2}_{r}} \mathbf{K}_{0}[k]
  +\int_{\Sigma^{3}_{r}} \mathbf{Z}[k]\,,
\end{equation}

\noindent
which is the gravitational analog of Eq.~(\ref{eq:QversusJ}). In Section~\ref{bardeen} we will show how the integral of
$\mathbf{K}_{0}[k]$ over $S^{2}_{r}$ goes to zero while the volume integral
goes to $M/2$ in the $r\to 0$ limit for the regular magnetic\footnote{For
  regular dyonic black hole solutions see \cite{Bokulic:2025brf}.} Bardeen
solution .

This paper is organized as follows: in Section~\ref{sectionaction} we study
the action and the equations of motion for a generic Einstein--NLED theory
within the Wald formalism. In Section~\ref{sectionsymmetry} we study the local
symmetries and conserved charges.  In Section~\ref{thermodynamics} we derive
the Smarr formula from the generalized Komar charge and the first law of black
holes thermodynamics.  Section 5 is devoted to test the Smarr formula across
different solutions.  We begin in Section~\ref{electric} with the electric
case, examining the Einstein--Born--Infeld and the electric ModMax black
holes. In Section~\ref{magnetic} we treat the Hayward and the magnetic ModMax
black holes. Finally, Section~\ref{bardeen} contains a detailed analysis of
the thermodynamics of the Bardeen solution.  Section~\ref{sectionconclusion}
discusses our results, open questions, and possible future
directions. Appendix \ref{Appendix} contains the test of the Smarr formula for
the magnetic Reissner--Nordstr\"om black hole which we can compare to the one
of regular black holes.

\section{Action for Einstein--NLED theories}\label{sectionaction}

In this paper we are going to use the conventions of
Ref.~\cite{Ortin:2015hya}, with a mostly minus signature, the Vielbein
$e^{a}=e^{a}{}_{\mu}dx^{\mu}$ as gravitational field and differential-form
notation.
	
In $d=4$, the most general Lorentz-symmetric Lagrangian for non-linear
electrodynamics (NLED) theories is built from a scalar and a pseudoscalar
\cite{Sorokin:2021tge}, here respectively denoted by $X$ and $Y$, defined as
	
\begin{equation}
		 X(e^{a},A)\equiv -\frac{1}{4} F_{\mu\nu}F^{\mu\nu}\,,		 \qquad
		 Y(e^{a},A)\equiv \frac{1}{4} F_{\mu\nu}(\star F)^{\mu\nu}\,,
\end{equation}

\noindent
where $F=\dd A $ is the 2-form field-strength and $\star$ denotes the Hodge
star operator.  The interaction Lagrangian depends on $X$ and $Y$ through a
coupling constant, here denoted by $\alpha$, and over a generic spacetime we
denote it as

\begin{equation}
		\mathbf{L}\equiv \dd ^{4} x  \sqrt{|g|}\mathscr{L}(X,Y,\alpha),
\end{equation}

\noindent
and its derivatives with respect to $X$ and $Y$  as
	
\begin{equation}
		\mathbf{L}_{X}= \dd ^{4} x
                \sqrt{|g|}\mathscr{L}_{X}(X,Y,\alpha),
                \qquad
		\mathbf{L}_{Y}= \dd ^{4} x  \sqrt{|g|}\mathscr{L}_{Y}(X,Y,\alpha).
\end{equation}
	
The action for a generic Einstein--NLED theory in differential-form language
the takes the form
	
\begin{equation}
  S[e^{a}, A, \alpha(x), C]
  =
  \frac{1}{16\pi G_{N}^{(4)}}\int\bigg\{-\star(e^{a}\wedge e^{b})\wedge R_{ab}
  +\mathbf{L}\big(e^{a},A,\alpha(x)\big)\bigg\}\,.
\end{equation}

In order to apply Wald's formalism, it is necessary (see for example
Ref.~\cite{Meessen:2022hcg}) to promote the coupling constant $\alpha$ to a
dynamical field $\alpha(x)$ by introducing an auxiliary $3$-form field $C$
whose equation of motion will guarantee the on-shell constancy of $\alpha$:
	
\begin{equation}
  \begin{aligned}
    \label{actionS}
    S[e^{a}, A, \alpha(x), C]
    &=
      \frac{1}{16\pi G_{N}^{(4)}}\int\bigg\{-\star(e^{a}\wedge e^{b})\wedge
      R_{ab}
      +\mathbf{L}\big(e^{a},A,\alpha(x)\big)+ \alpha(x) \dd C\bigg\}
      &                         		\\
    \\ & \equiv  \int \mathbf{L}^{TOT}\,.
  \end{aligned}
\end{equation}

The variation of the action Eq.~\eqref{actionS} is 

\begin{equation}
  \label{eq:generalvariationtheory}
		\delta S[e^{a}, A, \alpha(x), C]=
		\int \bigg\{
		\mathbf{E}_{a}\wedge \delta e^{a}
		+\mathbf{E}_{A} \wedge \delta A	
		+\mathbf{E}_{\alpha}\delta \alpha
		+\mathbf{E}_{C} \wedge\delta C
		+\dd \mathbf{\Theta}
		\bigg\}\,,
\end{equation}

\noindent
where the equations of motion and the presymplectic potential are
	
\begin{align}
		\mathbf{E}_{a} &= \frac{1}{16\pi G^{(4)}_{N}}
		\bigg[
		\imath_{a} \star(e^{b}\wedge e^{c})\wedge R_{bc}
		-\imath_{a} \mathbf{L}+
		\imath_{a} F \wedge \star \Pi
		\bigg]\,,\\\nonumber\\
		\mathbf{E}_{A} &=\frac{1}{16\pi G^{(4)}_{N}}\bigg[-\dd \star \Pi
		\bigg]\,,\\\nonumber\\
		\mathbf{E}_{\alpha}&=\frac{1}{16\pi G^{(4)}_{N}}\bigg[ \mathbf{L}_{\alpha} + \dd C
		\bigg]\,,\\
		\nonumber\\
		\mathbf{E}_{C}&=\frac{1}{16\pi G^{(4)}_{N}}\bigg[-\dd \alpha\bigg]\,,
		\\
		\nonumber\\
		\mathbf{\Theta}&=\frac{1}{16\pi G^{(4)}_{N}}\bigg[-\star(e^{a}\wedge e^{b})\wedge \delta \omega_{ab}+\star \Pi\wedge \delta A+\alpha\delta C\bigg]	\,,
\end{align}

\noindent
where we have defined 
	
\begin{equation}
  \label{pi}
		\Pi\equiv \mathscr{L}_{X} F -\mathscr{L}_{Y} \star F\qquad\qquad
		\star\Pi= \mathscr{L}_{X} \star F +\mathscr{L}_{Y} F\,.
\end{equation}
	
As we can see from the expression of $\mathbf{E}_{C}$, $\alpha$ is on-shell
constant. The equation of motion for $\alpha$ implies, on shell, that the
auxiliary field $C$ is fixed by the other dynamical fields and does not carry
extra degrees of freedom. The extended action \eqref{actionS} is therefore
equivalent, on shell, to the original theory with \(\alpha\) a constant.

For later convenience we write the Einstein and non-linear Maxwell equations
in components
	
\begin{alignat}{2}
  \label{EG}
  \mathbf{E}_{a} \doteq 0 \qquad&\Rightarrow\qquad
		&\; G_{\mu\nu} \doteq \tfrac{1}{2}\Big[(\mathscr{L}-\mathscr{L}_{Y}\,Y)\,g_{\mu\nu}
		+ \mathscr{L}_{X}\,F_{\mu}{}^{\sigma}F_{\nu\sigma}\Big], \\[6pt]
		\mathbf{E}_{A} \doteq 0 \qquad&\Rightarrow\qquad
		&\; \nabla_{\mu}\!\bigl(\mathscr{L}_{X}\,F^{\mu\nu}-\mathscr{L}_{Y}\,(\star F)^{\mu\nu}\bigr)
		\doteq 0.\label{EL}
\end{alignat}
	
From now on, we will denote $H \equiv \dd C$ for simplicity. In the next
section we will study the local (Lorentz, gauge and diffeomorphism) symmetries
of this theory and compute the corresponding Noether currents and conserved
charges.

\section{Symmetries and conserved charges}\label{sectionsymmetry}

\subsection{Lorentz Symmetry}

The action Eq.~\eqref{actionS} is exactly invariant under local
Lorentz transformations which only act on the Vielbein

\begin{equation}
	\delta_{\sigma}e^{a}
	=
	\sigma^{a}{}_{b}e^{b}\,,
\end{equation}

\noindent
where $\sigma^{ab}$ is antisymmetric, and on the fields derived from it

\begin{align}
	\delta_{\sigma}\omega^{ab}
	&=
	\mathcal{D}\sigma^{ab}\,,		
	\\
	\nonumber
	\\
	\delta_{\sigma}R^{ab}
	&=
	-2\sigma^{[a}{}_{c}R^{b]c}\,.
\end{align}

Particularizing Eq.~(\ref{eq:generalvariationtheory}) to this symmetry we
find the Noether identity

\begin{equation}
	\label{eq:LorentzNoetheridentity}
	\mathbf{E}^{[a}\wedge e^{b]}
	=
	0\,,
\end{equation}

\noindent
which asserts the symmetry of the energy-momentum tensor, and the closed
$3$-form current

\begin{equation}
	\mathbf{J}[\sigma]
	=
	\frac{-1}{16\pi G_{N}^{(4)}} \star (e^{a}\wedge e^{b})\wedge \mathcal{D}\sigma_{ab}
	=
	\dd\mathbf{Q}_{L}[\sigma]\,,
\end{equation}

\noindent
with the Noether $2$-form charge

\begin{equation}
	\label{eq:LorentzNoetherchargeCS}
	\mathbf{Q}_{L}[\sigma]
	=
	\frac{-1}{16\pi G_{N}^{(4)}} \star (e^{a}\wedge e^{b})\sigma_{ab}\,.
\end{equation}

As we have stressed elsewhere (see Ref. ~\cite{Barbagallo:2025qdy} for
instance), the Lorentz charge is closed on-shell for covariantly constant
Lorentz parameters

\begin{equation}
  \mathcal{D}\sigma^{ab}=0 \qquad \Rightarrow \qquad
  \dd  	\mathbf{Q}_{L}[\sigma]=0.
\end{equation}
       
Note that this condition is equivalent to requiring that the transformations
leave the spin connection invariant.  It is fortunate that the invariance of
the Vielbein is not required, as it is impossible to fulfill this condition.

In the end, notice that for the \textit{Killing bivector}

\begin{equation}
	\label{eq:Killingbivectordef}
	P_{k}{}^{ab}
	\equiv
	\nabla^{a}k^{b}\,,
\end{equation}

\noindent
where $k$ is a Killing vector, $\mathbf{Q}_{L}[P_{k}]$ is the standard Komar
$2$-form \cite{Komar:1958wp} which is on-shell closed in pure gravity.

\subsection{$C$ field gauge symmetry}

The invariance of the action under the gauge symmetry
	
\begin{equation}
\delta_{\chi^{H}} C = \dd \chi^{H}, \qquad \chi^{H} \in \Omega^{(2)}\,,
\end{equation}

\noindent
implies the   off-shell identity
	
\begin{equation}\label{dec}
		{\dd \mathbf{E}_{C}=0}\,,
\end{equation}

\noindent
and 
	
\begin{equation}
  \mathbf{J}_{C}(\chi) = \dd \mathbf{Q}_{C}(\chi)\,,
  \qquad\qquad
\mathbf{Q}_{C}(\chi)= \frac{1}{16\pi G_{N}^{(4)}} \alpha \chi.
\end{equation}

Notice that if $\chi = k$ is a Killing parameter generating transformations
that annihilate all the fields (just $C$ in this case), this charge is
on-shell closed
	
\begin{equation}
		{\dd \mathbf{Q}_{C} (k)\doteq0}.
\end{equation}

\subsection{$A$ field gauge symmetry}

The invariance of the action under the gauge symmetry\footnote{We don't have
  an explicit form for $\mathbf{L}$, nonetheless we suppose it is exactly
  gauge-invariant.}
	
\begin{equation}
		\delta_\chi A = \dd \chi, \qquad \chi \in \Omega^{(0)}\,,
\end{equation}

\noindent
implies the  off-shell identity
	
\begin{equation}\label{dea}
		{\dd \mathbf{E}_{A}=0}\,,
\end{equation}

\noindent
and 
	
\begin{equation}
\mathbf{J}_{A}(\chi) = \dd \mathbf{Q}_{A}(\chi)\,,\qquad\qquad
\mathbf{Q}_{A}(\chi)=	\frac{1}{16\pi G_{N}^{(4)}}(\chi \star \Pi)\,,
\end{equation}

\noindent
where $\Pi$ is defined in Eq.~\eqref{pi}.

If $\chi = k$ is a Killing parameter, this charge is on-shell closed, as
expected
	
\begin{equation}
{\dd \mathbf{Q}_{A} (k)\doteq0}.
\end{equation}

Choosing the Killing parameter $k = 1$ we can define a electric charge 
	
\begin{equation}\label{elecharge}
  {\mathbf{{Q}}=\frac{1}{	16 \pi G_{N}^{(4)}}\star \Pi,
    \hspace{2 cm}
\mathcal{Q} \equiv \int_{S^{2}}\mathbf{{Q}}}\,,
\end{equation}

\noindent
satisfying a Gauss law on-shell. We can also define a magnetic charge
as\footnote{Notice that, if we can write $F = \dd A$ globally, this charge is
  automatically a total derivative and its integral over closed 2-dimensional
  surfaces will vanish.}
	
\begin{equation}\label{magncharge}
\mathbf{P}=\frac{1}{16 \pi G_{N}^{(4)}} F,
    \hspace{2 cm}
\mathcal{P} \equiv \int_{S^{2}}\mathbf{{P}}\,,
\end{equation}

\noindent
which satisfies a Gauss law by virtue of the Bianchi identity.

\subsection{GCTs invariance}

\subsubsection{Noether-Wald charge}

When acting on fields with gauge freedoms, GCTs induce gauge transformations
(see Refs.~\cite{Elgood:2020svt,Elgood:2020nls}).  Therefore, given a vector
field $\xi$, its infinitesimal GCT, $\delta_{\xi}$, will be given by the usual
Lie derivative along $\xi$ plus an \textit{induced} or \textit{compensating}
gauge transformations with gauge parameters, here generically denoted by
$\Lambda_{\xi}$, that depend on the fields as well as on $\xi$

\begin{equation}
	\delta_{\xi}  = -\mathcal{L}_{\xi} + \delta_{\Lambda_{\xi}}.
\end{equation}
      
These gauge parameters are fully determined when $\xi$ generates a symmetry of
all the fields and $\delta_{\xi}$ annihilates them all. In that case, we
denote it as $\xi=k$ since the vector field that generates the symmetry must,
in particular, be a Killing vector of the metric, whose gauge freedoms are
only GCTs.

The parameters of the compensating local Lorentz transformations are given by 

\begin{equation}
	\label{eq:sigmakdef}
	\sigma_{k}{}^{a}{}_{b}
	=
	\imath_{k}\omega^{a}{}_{b} -P_{k}{}^{a}{}_{b}\,,
\end{equation}

\noindent
where $P_{k}{}^{a}{}_{b}$ is the \textit{Lorentz momentum map}, defined in
Eq.~(\ref{eq:Killingbivectordef}).

The parameters of the compensating gauge transformation for the gauge field
$A$ are given by

\begin{equation}
	\label{eq:Lambdak}
	\chi_{k}
	=
	\imath_{k}A -P_{k}\,,
\end{equation}

\noindent
where $P_{k}$ is the $0$-\textit{form} \textit{momentum map} defined by the
momentum map equation

\begin{equation}
	\label{eq:pformmomentummap}
	\imath_{k}F+\dd P_{k} 
	=
	0\,.
\end{equation}

The parameters of the compensating gauge transformation for the gauge field
$C$ are given by

\begin{equation}
	\label{eq:cLambdak}
	\chi_{k}^{H}
	=
	\imath_{k}C -P^{H}_{k}\,,
\end{equation}

\noindent
where $P^{H}_{k}$ is the $2$-\textit{form} \textit{momentum map} defined by
the momentum map equation

\begin{equation}
\label{eq:cformmomentummap}
\imath_{k}H+\dd P^{H}_{k} 
=
0\,.
\end{equation}

For later convenience, notice that, since, by assumption, the GCT generated by
$k$ annihilates all fields and since $\delta_{\Lambda_{\xi}}\star \Pi=0$, we have

\begin{equation}
  0
  =
  \delta_{k}\star \Pi
  =
  - \mathcal{L}_{k} \star\Pi
  =
  -\dd\imath_{k} \star\Pi-\imath_{k} \dd \star\Pi\doteq -\dd\imath_{k} \star\Pi\,.
\end{equation}

\noindent
Therefore, the must exist a function $\tilde{P}_{k}$ such that

\begin{equation}
	{\imath_{k} \star\Pi +\dd \tilde{P}_{k}\doteq 0}\, ,
\end{equation}
      
this is the \textit{dual-momentum map equation}.

Then, the $\delta_{\xi}$ transformations take the general form 

\begin{equation}
\label{eq:deltaxigeneral}
\delta_{\xi}
\equiv
-\mathcal{L}_{\xi} +\delta_{\sigma_{\xi}} +\delta_{\chi_{\xi}} +\delta_{\chi^{H}_{\xi}}\,,
\end{equation}

\noindent
and their action on all the fields of this theory

\begin{subequations}
\begin{align}
	\delta_{\xi}e^{a}
		& =
		-\left(\mathcal{D}\xi^{a}+P_{\xi}{}^{a}{}_{b}e^{b}\right)\,,
		\\
		& \nonumber \\
		\delta_{\xi}A
		& =
		-\left(\imath_{\xi}F+\dd P_{\xi}\right)\,,
		\\
		& \nonumber \\
		\delta_{\xi}C
		& =
		-\left(\imath_{\xi}H+\dd P^{H}_{\xi}\right)\,,
		\\
		& \nonumber \\
		\delta_{\xi}\alpha
		& =
		-\imath_{\xi} \dd \alpha\,,
	\end{align}
\end{subequations}

\noindent
is guaranteed to vanish when $\xi=k$.

Let us now derive the Noether charge associated to GCTs, also known as the
\textit{Noether--Wald} charge. Under the transformations $\delta_{\xi}$, the
action is invariant up to a total derivative:

\begin{equation}
	\delta_{\xi}S
	=
	-\int \dd \imath_{\xi}\mathbf{L}^{TOT}\,.
\end{equation}
      
On the other hand, particularizing Eq.~\eqref{eq:generalvariationtheory} to
the $\delta_{\xi}$ transformation, integrating by parts and using the Noether
identities Eqs.~\eqref{dec}, \eqref{dea} and
(\ref{eq:LorentzNoetheridentity}), we get 

\begin{equation}\label{gcts}
	\int \bigg\{-\bigg[\mathcal{D} \mathbf{E}_{a} +  \mathbf{E}_{A} \wedge \imath_{a} F 
	+  \mathbf{E}_{C} \wedge \imath_{a} H
	+  \mathbf{E}_{\alpha} \imath_{a} \dd \alpha \bigg]\xi^{a}+ \dd \mathbf{J}_{NW}(\xi) \bigg\}=0\,,
\end{equation}

\noindent
where

\begin{equation}\label{Jgct}
	\mathbf{J}_{NW}(\xi)= \mathbf{\Theta} +   \mathbf{E}_{a} \xi^{a} +  \mathbf{E}_{A}   P_{\xi}
	+\mathbf{E}_{C} \wedge P_{\xi}^{H} +\imath_{\xi} \mathbf{L}^{TOT}.
\end{equation}

Eq.~\eqref{gcts}  implies the following off-shell identities
	
\begin{equation}
\begin{aligned}
\mathcal{D}\mathbf{E}_{a} +  \mathbf{E}_{A} \wedge \imath_{a} F 
+  \mathbf{E}_{C} \wedge \imath_{a} H
+  \mathbf{E}_{\alpha} \imath_{a} \dd \alpha&=0\,,\\
& \\
			 \dd \mathbf{J}_{NW}&=0\,,
\end{aligned}
\end{equation}

\noindent
the first of which is a Noether identity which automatically satisfied and the
second of which implies that  

\begin{equation}
\mathbf{J}_{NW}(\xi) = 	\dd\mathbf{Q}_{NW}(\xi)\,,
\end{equation}

\noindent
where 
	
\begin{equation}
  \mathbf{Q}_{NW}(\xi)
  =
  \frac{1}{16\pi G_{N}^{(4)}}  \bigg\{\star (e^{a} \wedge e^{b})P_{\xi ab}
  -\star \Pi P_{\xi} -\alpha P^{H}_{\xi} \bigg\}\,,
\end{equation}

\noindent
is the Noether--Wald charge of the theory.

\subsubsection{Generalized Komar charge}

From Eq.~\eqref{Jgct} we have

\begin{equation}
\dd \mathbf{Q}_{NW}[\xi]
=
\mathbf{J}_{NW}(\xi)
=
\mathbf{\Theta} +\mathbf{E}_{a} \xi^{a}
+\mathbf{E}_{A} P_{\xi}	+\mathbf{E}_{C} \wedge P_{\xi}^{H}
+\imath_{\xi} \mathbf{L}^{TOT}\,,
\end{equation}

\noindent
which in general does not vanish. The first term in the right-hand side vanishes for
$\xi=k$. The second, third and fourth terms vanish on-shell and we are left
with

\begin{equation}
	\label{eq:dQk}
	\dd \mathbf{Q}_{NW}[k]
	\doteq 
	\imath_{k}\mathbf{L}^{TOT}\,.  
\end{equation}

This equation, which expresses the failure of the Noether-Wald charge to be a
closed $2$-form, can be directly used to derive Smarr formulas
\cite{Smarr:1972kt}, following Ref.~\cite{Liberati:2015xcp}. However, as
discussed in Refs.~\cite{Ortin:2021ade,Cerdeira:2025elp},\footnote{See also
  Ref.~\cite{Golshani:2024fry}, based on the results of
  Ref.~\cite{Adami:2024gdx}.} since the left-hand side of the identity is a total
derivative the sum of these two terms must also be a total derivative. If we
manage to rewrite the right-hand side as a total derivative different from
$\dd\mathbf{Q}_{NW}[k]$, call it $\dd \boldsymbol{\omega}_{k}$, then

\begin{equation}
\label{eq:generalconstructiongeneralizedKomarcharge}
\mathbf{K}[k]
\equiv
\boldsymbol{\omega}_{k}-\mathbf{Q}_{NW}[k]\,,
\end{equation}

\noindent
will be an on-shell closed $2$-form charge, called \textit{generalized Komar
  charge}. The derivation of the Smarr formula from an on-shell closed
generalized Komar charge along the lines of Refs.~\cite{Kastor:2008xb,
  Bardeen:1973gs,Carter:1973rla,Magnon:1985sc,Bazanski:1990qd,Kastor:2010gq}
is simpler and physically more transparent. The procedure outlined above can
be used to derive systematically generalized Komar charges for many theories
(see,
\textit{e.g.}~Refs.~\cite{Mitsios:2021zrn,Meessen:2022hcg,Ortin:2022uxa}).

How can we find a $\boldsymbol{\omega}_{k}\neq \mathbf{Q}_{NW}[k]$? As
explained in Ref.~\cite{Ortin:2024mmg}, Eq.~(\ref{eq:dQk}) has been obtained
by computing $\imath_{k}\mathbf{L}^{TOT}$ first and then using the equations
of motion. In this order, these operations give
$\dd\mathbf{Q}_{NW}[k]$. However, in the reverse order, they give a formally
different expression, $\dd\boldsymbol{\omega}_{k}$, which, on-shell, should be
identical. The difference between $\dd\boldsymbol{\omega}_{k}$ and
$\dd\mathbf{Q}_{NW}[k]$, which we have called $\dd\mathbf{K}[k]$, must vanish
on-shell by construction, which proves that $\mathbf{K}[k]$ in
Eq.~(\ref{eq:generalconstructiongeneralizedKomarcharge}) is on-shell closed.

The calculation of $\mathbf{L}^{TOT}$ on-shell and of
$\boldsymbol{\omega}_{k}$ is greatly simplified when there is a global
transformation of the fields of the theory that rescales $\mathbf{L}^{TOT}$
\cite{Cerdeira:2025elp}. Performing a generic rescaling

\begin{equation}
	e^{'a}= \lambda e^{a},\qquad
	A'= \lambda^{\beta} A,\qquad
	\alpha'= \lambda^{\gamma}\alpha,\qquad
	C'= \lambda^{\delta} C\,,
\end{equation}

\noindent
in the action \eqref{actionS} the Einstein--Hilbert term scales as
$\lambda^{2}$: the vielbeins transform, whereas $R_{ab}$ remains invariant
because the spin connection does not. Consequently, the other two terms in the
action must also scale as $\lambda^{2}$. We set $\beta=1$ because we restrict
ourselves to Einstein--NLED theories that reduces to Einstein-Maxwell theory
(for which $\beta=1$) in a suitable limit of the coupling constant
$\alpha$. We leave $\gamma$ free, because it depends on the particular
Einstein--NLED theory considered. From the last term in the action, we deduce
that $\delta = \gamma-2$.


Hence, a generic Einstein--NLED Lagrangian satisfies the following
quasi-homogeneity condition

\begin{equation}\label{Ltot}
		{\mathbf{L}^{TOT}[\lambda e^{a},\lambda A,\lambda F,
                  \lambda^{\gamma}\alpha,\lambda^{2-\gamma}
                  C,\lambda^{2-\gamma} H]
                  =
                  \lambda^{2}\mathbf{L}^{TOT}[e^{a},A,F,\alpha,C,H]}\,,
\end{equation}

\noindent
where we assumed the fields $A,C$ and their field strengths $F,H$ are
independent variables and rescale in the same way.

Deriving Eq.~\eqref{Ltot} with respect to $\lambda$ using the chain rule and
then setting $\lambda = 1$ we get

\begin{equation}
  \begin{aligned}
    &\mathbf{E}_{a}\wedge  e^{a} 
      +\mathbf{E}_{A}\wedge A
      +\gamma
      \mathbf{E}_{\alpha} \alpha 
      +(2-\gamma)\mathbf{E}_{C}\wedge C
      +
      \frac{1}{16\pi G_{N}^{(4)}}\dd  
      \bigg(\star \Pi\wedge A +(2-\gamma)\alpha C\bigg)
    \\
    & =
      2 \mathbf{L}^{TOT}\,,
  \end{aligned}
\end{equation}

\noindent
and on-shell
	
\begin{equation}
  \label{Ltotonshell}
		\mathbf{L}^{TOT} \doteq \frac{1}{16\pi G_{N}^{(4)}}
		\dd  (\tfrac{1}{2}\star \Pi\wedge A +\frac{(2-\gamma)}{2}
		\alpha C)\equiv \dd \Xi.
\end{equation}

By assumption, $\delta_{k}\Xi=0$. Since $\Xi$ is not gauge invariant
	
\begin{align*}
		0 = \delta_{k} \Xi &=-\mathcal{L}_{k}\Xi +\delta_{\chi_{k}}\Xi
		\doteq
		- (\dd\imath_{k}+\imath_{k} \dd)\Xi 
		+\frac{1}{16\pi G_{N}^{(4)}}\dd \bigg(\tfrac{1}{2}\star \Pi  \chi_{k} +\frac{(2-\gamma)}{2}
		\alpha   \chi_{k}^{H}\bigg).
\end{align*} 

The last equation can be used to write $\imath_{k} \mathbf{L}^{TOT}$ as a
total derivative, and
	
\begin{equation}
  \dd \mathbf{Q}_{NW}(k)
  \doteq
  \imath_{k} \mathbf{L}^{TOT}
  \doteq
  \frac{1}{16\pi G^{(4)}_{N}}
		\dd\bigg\{
		-\tfrac{1}{2}  P_{k} \star \Pi
		-	\tfrac{1}{2}\tilde{P}_{k}  F
		-\frac{(2-\gamma)}{2}\alpha P_{k}^{H}\bigg\}
\equiv  \dd \omega_{k}\,.
\end{equation}

The generalized Komar charge is finally\footnote{
		Note that for the Maxwell theory
		
		\begin{equation}
			\mathbf{L}= \dd ^{4} x \sqrt{|g|}X, \qquad \mathscr{L}_{X}=1, \quad \mathscr{L}_{Y}=0
		\end{equation}
		we get the well-known result \cite{Ortin:2022uxa}
		
		\begin{equation}
			\mathbf{K}(k)=
		\frac{1}{16\pi G_{N}^{(4)}}
		\bigg\{-\star (e^{a} \wedge e^{b})P_{k ab}
		+\tfrac{1}{2}  P_{k} \star F
		-	\tfrac{1}{2}\tilde{P}_{k}  F		
		\bigg\}\,.
		\end{equation}
              }
	
\begin{equation}
\label{komar}
\mathbf{K}[k]
=
\frac{1}{16\pi G_{N}^{(4)}}
\bigg\{-\star (e^{a} \wedge e^{b})P_{k ab}
+\tfrac{1}{2}  P_{k} \star \Pi
-	\tfrac{1}{2}\tilde{P}_{k}  F
+\frac{\gamma}{2}\alpha P_{k}^{H}
\bigg\}\,,
\end{equation}

\noindent
which can be easily verified to be on-shell closed by direct computation using
the assumptions made in its derivation
	
\begin{equation}
  \dd \mathbf{K}[k]\doteq 0\,.
\end{equation}

\section{Black hole thermodynamics}\label{thermodynamics}

\subsection{Smarr Formula}

The event horizon of asymptotically-flat, stationary black holes is the
Killing horizon of a Killing vector that, in adapted coordinate, can be
written as

\begin{equation}
	k = \partial_{t} -\Omega \partial_{\varphi}\,,
\end{equation}

\noindent
where $\varphi$ is the angular coordinate around the axis of rotation and
Ω is the angular velocity of the horizon. Assuming the existence of a
bifurcation surface $\mathcal{BH}$, we can choose a spacelike hypersurface
$\Sigma^{3}$ with boundary
$\partial \Sigma^{3} =S^{2}_{\mathcal{BH}}\cup S^{2}_\infty$ and, integrating
$\dd \mathbf{K}[k]\doteq 0$ over $\Sigma^{3}$ and applying Stokes theorem

\begin{equation}
  \int_{S^{2}_{\mathcal{BH}}} \mathbf{K}[k]
  \doteq   
	\int_{S^{2}_{\infty}} \mathbf{K}[k]\,.
\end{equation}

\paragraph{Integral at infinity}

The gravitational term gives 
	
\begin{equation}
  -\frac{1}{16\pi G_{N}^{(4)}}
  \int_{S^{2}_\infty}   \star (e^{a} \wedge e^{b})P_{k ab} 
  =
  \tfrac{1}{2} M -\Omega J,
\end{equation}

\noindent
where $J$ is the angular momentum.
	
If $F$ vanishes asymptotically
($ F\stackrel{r\rightarrow\infty}{\rightarrow}0$), we have

\begin{align}\label{pinf1}
\imath_{k} F &\quad\stackrel{r\rightarrow\infty}{\rightarrow}0
&\Rightarrow\quad \dd P_{k} &\quad\stackrel{r\rightarrow\infty}{\rightarrow}0
		&\Rightarrow\quad P_{k} &\quad\stackrel{r\rightarrow\infty}{\rightarrow}\mathrm{constant}\equiv\Phi_{\infty},
		\\\nonumber\\\label{pinf2}
		\imath_{k}\star\Pi &\quad\stackrel{r\rightarrow\infty}{\rightarrow}0
		&\Rightarrow\quad \dd\tilde{P}_{k} &\quad\stackrel{r\rightarrow\infty}{\rightarrow}0
		&\Rightarrow\quad \tilde{P}_{k} &\quad\stackrel{r\rightarrow\infty}{\rightarrow}\mathrm{constant}\equiv\Psi_{\infty}\,,
\end{align}

\noindent
where $\Phi_{\infty}$ and $\Psi_{\infty}$ are, respectively, the electrostatic
and magnetostatic potentials at infinity.

If $F$ does not vanish asymptotically, then $P_{k}$ and $\tilde{P}_{k}$ are
not closed anymore and, for the Hodge theorem, they can be decomposed into an
harmonic component (which are constants for $0$-forms) and a coexact component
	
\begin{equation}
  P_{k} = \dd^\dag \zeta + \mathrm{constant}, \qquad 
  \tilde{P}_{k} = \dd^\dag \tilde\zeta + \mathrm{constant}, \qquad \zeta, \tilde \zeta \in \Omega^{(1)}.
\end{equation}
              
However, as proven in \cite{Zatti:2024vqv}, only the closed part of
$\tilde{P}_{k}$ can contribute to the above integral.  So we get
	
\begin{subequations}
  \begin{align}
    \frac{1}{16\pi G_{N}^{(4)}}\int_{S^{2}_\infty} 
    \tfrac{1}{2}  P_{k} \star \Pi
    & =
      \tfrac{1}{2}\Phi_{\infty} \mathcal{Q}\,,
    \\
    & \nonumber \\
    -\frac{1}{16\pi G_{N}^{(4)}}	\int_{S^{2}_\infty} \tfrac{1}{2}F  \tilde{P}_{k}
    & \doteq
      -\tfrac{1}{2}\Psi_\infty \mathcal{P}\,.
  \end{align}
\end{subequations}
	
Defining 
	
\begin{equation}
{\Lambda\equiv\frac{1}{16\pi G_{N}^{(4)}}	\int_{S^{2}}   {P}^{H}_{k}}\,,
\end{equation}

\noindent
the last term in Eq.~\eqref{komar} is $\frac{\gamma}{2} \alpha \Lambda_\infty.$

\paragraph{Integral over the bifurcation sphere}

The gravitational term gives 
	
\begin{equation}
  -\frac{1}{16\pi G_{N}^{(4)}}	\int_{S^{2}_{\mathcal{BH}}} 
  \star (e^{a} \wedge e^{b})P_{k ab} = \frac{\kappa}{8\pi G_{N}^{(4)}} A_{\mathcal{H}}= TS\,,
\end{equation}

\noindent
where $P_{k}^{ab}\stackrel{\mathcal{BH}}{=}\kappa n^{ab}$, with $n^{ab}$ the
binormal to the horizon, $\kappa$ is the surface gravity of the horizon,
$A_{\mathcal{H}}$ is its area, $T = \frac{\kappa}{2\pi}$ is the Hawking
temperature and $S = \frac{A_{\mathcal{H}}}{4 G_{N}^{(4)}}$ is its
Bekenstein-Hawking entropy.
	
On the bifurcation surface $\mathcal{BH}$ we have by definition
$k\stackrel{\mathcal{BH}}{=}0$, thus,
	
\begin{align}
\imath_{k} F &\stackrel{\mathcal{BH}}{=}0
		&\Rightarrow\quad \dd P_{k} &\stackrel{\mathcal{BH}}{=}0
		&\Rightarrow\quad P_{k} &\stackrel{\mathcal{BH}}{=}\mathrm{constant}\equiv\Phi_{\mathcal{BH}},\label{pbh1}
		\\\nonumber\\
		\imath_{k}\star\Pi &\stackrel{\mathcal{BH}}{=}0
		&\Rightarrow\quad \dd\tilde{P}_{k} &\stackrel{\mathcal{BH}}{=}0
		&\Rightarrow\quad \tilde{P}_{k} &\stackrel{\mathcal{BH}}{=}\mathrm{constant}\equiv\Psi_{\mathcal{BH}}\,,\label{pbh2}
\end{align}

\noindent
where $\Phi_{\mathcal{BH}}$ and $\Psi_{\mathcal{BH}}$ are, respectively, the
electrostatic and magnetostatic potential at the bifurcation sphere. Therefore
we get
	
\begin{subequations}
  \begin{align}
    \frac{1}{16\pi G_{N}^{(4)}}\int_{S^{2}_{\mathcal{BH}}}
    \tfrac{1}{2}  P_{k} \star \Pi
    & =
      \tfrac{1}{2}\Phi_{\mathcal{BH}} \mathcal{Q}\,,
    \\
    & \nonumber \\
    -\frac{1}{16\pi G_{N}^{(4)}}	\int_{S^{2}_{\mathcal{BH}}} \tfrac{1}{2} F  \tilde{P}_{k}
    & \doteq 
    -\tfrac{1}{2}\Psi_{\mathcal{BH}}\mathcal{P}\,.
  \end{align}
\end{subequations}
	
The last term in Eq.~\eqref{komar} is
$ \frac{\gamma}{2} \alpha \Lambda_{\mathcal{BH}}\,.$
	
\paragraph{Smarr formula}

Combining these results we get 
	
\begin{equation}\label{smarrrrrr}
  M \doteq 2(TS+\Omega J )
    +(\Phi_{\mathcal{BH}}-\Phi_{\infty})\mathcal{Q}-(\Psi_{\mathcal{BH}}-\Psi_{\infty})\mathcal{P}	
+\gamma(\Lambda_{\mathcal{BH}}-\Lambda_{\infty})\alpha
\end{equation}

\noindent
that is the Smarr formula for asymptotically-flat, stationary black holes
solutions of Einstein--NLED theories. So far, this result has been obtained in
the literature (see
Ref. \cite{Breton:2004qa,Gulin:2017ycu,Balart:2017dzt,Zhang:2016ilt}) by
employing Euler's theorem for homogeneous functions. Using the Komar charge,
here we presented a general derivation of the Smarr formula that holds for any
Einstein--NLED theory and does not rely on homogeneity arguments.

\subsection{First  law}\label{sec-firstlawminimal5dsugra}

The first law of black-hole mechanics can be obtained by integrating the
exterior derivative of some on-shell-closed $2$-form $\mathbf{W}[k]$ over the
same spacelike hypersurface we used in the derivation of the Smarr
formula. Here we are going to review the construction of $\mathbf{W}[k]$ in
detail.

Denoting by $\varphi$ all the fields of the theory, the symplectic $3$-form is
defined as
	
\begin{equation}
  \label{omega}
  \omega(\varphi,\delta_{1}\varphi,\delta_{2}\varphi)
  \equiv
  \delta_{1}\mathbf{\Theta}(\varphi,\delta_{2}\varphi)
  -\delta_{2}\mathbf{\Theta}(\varphi,\delta_{1}\varphi)
  -\mathbf{\Theta}(\varphi,[\delta_{1},\delta_{2}]\varphi)\,,
\end{equation}
	
\noindent
where $\delta_{1,2}$ are arbitrary variations. In the following we choose
	
\begin{itemize}
\item$\delta_{1}=\delta$ (arbitrary, infinitesimal variations of the fields)
\item $\delta_{2}=\delta_{\xi}$ (infinitesimal GCTs of the fields, which
  include induced gauge transformations).
\end{itemize}	

In most of the literature, the last term in Eq.~\eqref{omega} has been ignored
because, if all the fields $\varphi$ are world tensors,
$\delta_{\xi}= -\mathcal{L}_{\xi}$ and
	
\begin{equation}
[\delta,\mathcal{L}_{\xi}]=0\,.
\end{equation}
              
When the fields have some gauge freedoms,
$\delta_{\xi}= -\mathcal{L}_{\xi}+\delta_{\Lambda_{\xi}}$ and (using
$[\delta, \dd]=0$)
	
\begin{equation}
\label{eq:commutatorofdeltas}
[\delta,\delta_{\xi}]
=
[\delta,\delta_{\Lambda_{\xi}}]
=
\delta_{\delta\Lambda_{\xi}}\,,
\end{equation}
	
\noindent
\textit{i.e.}~a gauge transformation with parameter $\delta\Lambda_{\xi}$. Our
theory has several gauge symmetries and we have to consider all of them.
	
Thus, our starting point is 
	
\begin{equation}
\begin{aligned}
  \omega(\varphi,\delta\varphi,\delta_{\xi}\varphi)
  &=
    \delta \mathbf{\Theta}(\varphi,\delta_{\xi}\varphi)
    -\delta_{\xi}\mathbf{\Theta}(\varphi,\delta\varphi)
    -\mathbf{\Theta}(\varphi,\delta_{\delta\Lambda_{\xi}}\varphi)
  \\
  \\
  & \doteq
    \delta\left[\dd\mathbf{Q}_{NW}[\xi]-\imath_{\xi}\mathbf{L}^{TOT}
    \right]
    -\left(-\mathcal{L}_{\xi}+\delta_{\Lambda_{\xi}}\right)
    \mathbf{\Theta}(\varphi,\delta\varphi)
    -\mathbf{\Theta}(\varphi,\delta_{\delta\Lambda_{\xi}}\varphi)\,,
\end{aligned}
\end{equation}

\noindent
where we used Eq.~(\ref{Jgct}) on-shell.  By definition of the Noether--Wald
charge and Lie derivative and using $[\delta,\dd ]=0$ and $\delta \xi =0$ in
the first step
	
\begin{equation}
\begin{aligned}
  \omega(\varphi,\delta\varphi,\delta_{\xi}\varphi)
  & \doteq
    \dd \delta\mathbf{Q}_{NW}[\xi]-\imath_{\xi}\delta \mathbf{L}^{TOT}
  \\
  &    \\
  & \hspace{.5cm}
    +\dd \imath_{\xi}\mathbf{\Theta}(\varphi,\delta\varphi)
    +\imath_{\xi}\dd \mathbf{\Theta}(\varphi,\delta\varphi)
    -\delta_{\Lambda_{\xi}} \mathbf{\Theta}(\varphi,\delta\varphi)
    -\mathbf{\Theta}(\varphi,\delta_{\delta\Lambda_{\xi}}\varphi)
  \\
  & \\
  & \doteq
    \dd \left[\delta\mathbf{Q}_{NW}[\xi]
    +\imath_{\xi}\mathbf{\Theta}(\varphi,\delta\varphi)\right]
    -\delta_{\Lambda_{\xi}} \mathbf{\Theta}(\varphi,\delta\varphi)
    -\mathbf{\Theta}(\varphi,\delta_{\delta\Lambda_{\xi}}\varphi)\,,
\end{aligned}
\end{equation}

\noindent
where in the last step we used
$\delta \mathbf{L}^{TOT}\doteq\dd \mathbf{\Theta}(\varphi,\delta\varphi)$.
Consider the last two terms of this expression. Using
	
\begin{align}
  \delta_{\sigma_{\xi}}\star (e^{a}\wedge e^{b})
  &=
    2\sigma_{\xi\;\;\; c}^{[a|}\star (e^{c}\wedge e^{|b]})\,,
  \\
  & \nonumber\\
  \label{deltaomega}
  \delta_{\sigma_{\xi}} \omega_{ab}
  &=
    \mathcal{D}\sigma_{\xi ab}\,,
  \\\nonumber\\
  \delta_{\sigma_{\xi}} \delta \omega_{ab}
  &=
    \delta\delta_{\sigma_{\xi}}  \omega_{ab}
    -\delta_{\delta \sigma_{\xi}} \omega_{ab}
    =
    \delta \mathcal{D} \sigma_{\xi ab}- \mathcal{D} \delta\sigma_{\xi ab}
    =
    -2\delta \omega_{[a|}{}^{c} \sigma_{\xi c |b]}\,,
\end{align}

\noindent
together with Eq.~(\ref{eq:commutatorofdeltas})
	
\begin{equation}\label{deltaVVV}
  \delta_{\chi_{\xi}}\delta A
  = ( \delta \delta_{\chi_{\xi}}-\delta_{\delta{\chi_{\xi}}})A
  =
  \delta \dd{\chi_{\xi}}- \dd \delta {\chi_{\xi}}
  =
  [\delta, \dd]{\chi_{\xi}}=0\,,
\end{equation}

\noindent
and, similarly, $\delta_{\chi^{H}_{\xi}} \delta C =0$, we get
	
\begin{equation}
  \label{deltath}
  \delta_{\Lambda_{\xi}}\mathbf{\Theta}(\varphi,\delta\varphi)=0
\end{equation}

\noindent
and
	
\begin{equation}
  -\mathbf{\Theta}(\varphi,\delta_{\delta \Lambda_{\xi}}\varphi)
  \doteq
  \frac{1}{16\pi G_{N}^{(4)}}\dd
  \bigg\{ \star (e^{a}\wedge e^{b}) 	\delta \sigma_{\xi ab}
  -\star \Pi   \delta \chi_{\xi}-\alpha \delta \chi^{H}_{\xi}\bigg\}\,.
\end{equation}

Combining these results we get
	
\begin{equation}
\label{eq:Wxidef}
\begin{aligned}
  \omega(\varphi,\delta\varphi,\delta_{\xi}\varphi)
  & \doteq
    \dd\bigg\{\delta\mathbf{Q}_{NW}[\xi]
    +\imath_{\xi}\mathbf{\Theta}(\varphi,\delta\varphi)
  \\
  &
    \hspace{1cm}+ \frac{1}{16\pi G_{N}^{(4)}}
  \bigg[\star ( e^{a}\wedge e^{b})\wedge  \delta \sigma_{\xi\, ab}
  -\star \Pi \delta \chi_{\xi}-\alpha\delta \chi^{H}_{\xi}\bigg]\bigg\}
  \\
  & \equiv \dd \mathbf{W}[\xi]\,.
\end{aligned}
\end{equation}

Notice that the symplectic potential vanishes if $\xi=k$, where
$\delta_{k}\varphi=0$ for all the fields of the theory. Our definitions of
$\delta_{\xi}$ taking into account the gauge freedoms of the fields ensure
that $\delta_{k}\varphi=0$ is a well-defined, gauge-independent
statement. Then, the 2-form $\mathbf{W}[k]$ is on-shell closed
	
\begin{equation}
  \dd\mathbf{W}[k] \doteq 0 \,.
\end{equation}

Using Eqs.~\eqref{eq:sigmakdef}, \eqref{eq:Lambdak}, \eqref{eq:cLambdak} and
expanding Eq.~\eqref{eq:Wxidef}, we get
	
\begin{equation}
\begin{aligned}
  \mathbf{W}[k]
  & \doteq\frac{1}{16\pi G_{N}^{(4)}}
    \bigg\{
    P_{k\, bc} \delta\star (e^{b}\wedge e^{c})
    -\imath_{k}\star (e^{a}\wedge e^{b})\wedge \delta \omega_{ab}
    -P_{k}   \delta  \star \Pi
    +\tilde{P}_{k} \delta F
    -P_{k}^{H}\delta \alpha
    \bigg\}\,,
\end{aligned}
\end{equation} 

\noindent
which is defined up to a total derivative.  Notice also that this expression
is manifestly invariant under gauge transformations $\delta_{\Lambda}$.
	
Integrating $\dd \mathbf{W}[k]\doteq0$ over the same timelike hypersurface we
used to derive the Smarr formula and applying Stokes theorem, we find
	
\begin{equation}
  \int_{S^{2}_{\infty}} \mathbf{W}[k]
  \doteq
  \int_{S^{2}_{\mathcal{BH}}} \mathbf{W}[k]\,.  
\end{equation}
	
\paragraph{Integral at infinity}

As we saw in Eqs.~\eqref{pinf1} and \eqref{pinf2}, the electric and magnetic
momentum maps are constant at infinity. Thus,
	
\begin{equation}
\begin{aligned}
\int_{S^{2}_{\infty}} \mathbf{W}[k]
& \doteq
\frac{1}{16\pi G_{N}^{(4)}}
\int_{S^{2}_{\infty}} \bigg\{  
P_{k\, bc} \delta\star (e^{b}\wedge e^{c})
-\imath_{k}\star (e^{a}\wedge e^{b})\wedge \delta \omega_{ab}
-P_{k}   \delta  \star \Pi
+\tilde{P}_{k} \delta F
-P_{k}^{H}\delta \alpha
\bigg\}
\\
& \\
& =
\delta M 
-\Omega\delta J
-\Phi_\infty \delta \mathcal{Q}
+\Psi_\infty \delta\mathcal{P}
-\Lambda_\infty \delta \alpha.
\end{aligned}
\end{equation}

\paragraph{Integral over the bifurcation surface}
Taking into account

$k \stackrel{\mathcal{BH}}{=} 0$, Eqs.~\eqref{pbh1}, \eqref{pbh2} plus the
general property $P_{k}^{ab}\stackrel{\mathcal{BH}}{=}\kappa n^{ab}$, with
normalization $n^{ab}n_{ab}=-2$, and the zeroth law satisfied by $\kappa$ we
have
	
\begin{equation}
\begin{aligned}
\int_{S^{2}_{\mathcal{BH}}} \mathbf{W}[k]
& \doteq
\frac{\kappa}{16\pi G_{N}^{(4)}}
\int_{S^{2}_{\mathcal{BH}}}n_{bc} \delta\star (e^{b}\wedge e^{c})
  +\frac{1}{16\pi G_{N}^{(4)}}\int_{S^{2}_{\mathcal{BH}}} \bigg\{
  -P_{k}   \delta  \star \Pi+\tilde{P}_{k} \delta F-P_{k}^{H}\delta \alpha
\bigg\}
\\
& \\
& =
\frac{\kappa \delta \mathcal{A}_{\mathcal{H}}}{8\pi G_{N}^{(4)}}
-\Phi_{\mathcal{BH}}  \delta \mathcal{Q} +\Psi_{\mathcal{BH}}\delta  \mathcal{P}
-\Lambda_{\mathcal{BH}} \delta \alpha.
\end{aligned}
\end{equation}

\paragraph{First law}
Combining these results we get
	
\begin{equation}
  \delta M
=
  T\delta S +\Omega\delta J
  +(\Phi_{\infty}-\Phi_{\mathcal{BH}})\delta \mathcal{Q} -(\Psi_{\infty}-\Psi_{\mathcal{BH}})\delta \mathcal{P}
  +(\Lambda_{\infty}-\Lambda_{\mathcal{BH}})\delta {\alpha}\,,
\end{equation}

\noindent
that is the first law for asymptotically-flat, stationary black holes
solutions of Einstein--NLED theories.\footnote{A different derivation can 
be found in Refs.~\cite{Ma:2014qma,Ma:2025dee}.}

\section{Testing the Smarr formula}

In this section we are going to consider some electric and magnetic black
holes solutions in order to verify the Smarr formula Eq.~\eqref{smarrrrrr}.

\subsection{Electric black holes}\label{electric}

\subsubsection{The ansatz}

Consider the following spherically metric
	
\begin{equation}
  \dd s^{2} = \lambda(r) \dd t^{2} - \frac{1}{\lambda(r) }\dd r^{2}
    -r^{2} (\dd \theta^{2} +\sin ^{2}\theta \dd \varphi^{2})\,,
\end{equation}

\noindent
with gauge potential
	
\begin{equation}
  \label{A}
A=A_{t}(r)\dd t \equiv a(r)\dd t\,,
\end{equation}

\noindent
that is a spherically symmetric non-linearly electric charged black hole. For
simplicity, we are going to assume that the Lagrangian $\mathscr{L}$ is just a
function of $X$ and does not depend on $Y$. Following the same line of
\cite{Fan:2016hvf, Toshmatov:2018cks}, for the non-linear Maxwell equations
\eqref{EL} we get
	
\begin{equation}
\label{nlmx}
\nabla_{\mu} ( \mathscr{L}_{X} F^{\mu\nu})\doteq 0
\qquad\Rightarrow \qquad
\frac2r+\frac{\mathscr{L}_{X}'}{\mathscr{L}_{X} }+ \frac{a''}{a'}\doteq 0\,,
\end{equation}

\noindent
where the prime denotes the  derivative with respect to $r$.
	
The Einstein equations are given by Eq.~\eqref{EG}
	
\begin{equation}
  G_{\mu\nu}\doteq  \tfrac{1}{2} \bigg(\mathscr{L}g_{\mu\nu}
  +\mathscr{L}_{X} F_{\mu\tau}F_{\nu\sigma}g^{\sigma \tau}\bigg)\,,
\end{equation}

\noindent
where the Einstein tensor is
	
\begin{equation}
  \label{matrix}
  (G_{\mu\nu})\doteq
  \left(
   \begin{array}{cccc}
			\displaystyle{\frac{\lambda  \left(r \lambda '+\lambda -1\right)}{r^{2}}} & 0 & 0 & 0 \\
			0 & \displaystyle{-\frac{r \lambda '+\lambda -1}{r^{2} \lambda } }& 0 & 0 \\
			0 & 0 & -\frac{1}{2} r \left(r \lambda ''+2 \lambda '\right) & 0 \\
			0 & 0 & 0 & -\frac{1}{2} r \sin ^{2}\theta  \left(r \lambda ''+2 \lambda '\right) \\
\end{array}
	\right)\,.
\end{equation}

There are only two  independent equations:
	
\begin{align}
  \frac{\lambda'}{r}+\frac{\lambda-1}{r^{2}}-\frac{\mathscr{L}}{2}+\tfrac{1}{2} \mathscr{L}_{X} a^{'2}&\doteq 0\,,\\
  \nonumber \\
  \lambda'' + \frac{2\lambda '}{r} -\mathscr{L}&\doteq 0\,.
\end{align}

Eq.~\eqref{nlmx}  is solved by 
	
\begin{equation}
  \label{LXQ}
  \mathscr{L}_{X}\doteq - \frac{4\mathcal{Q}G_{N}^{(4)}}{ r^{2} a'}.
\end{equation}

The normalization of the integration constant has been fixed so that, upon
integrating $\star\Pi \doteq \mathscr{L}_{X}\star F$ over a sphere, one
recovers the electric charge defined in Eq.~\eqref{elecharge}.
	
Choosing the following parametrization for the $\lambda$ function

\begin{equation}
\lambda(r;\alpha)\equiv 1-\frac{2m(r;\alpha)G_{N}^{(4)}}{r}\,,
\end{equation}

\noindent
the second Einstein equation gives
	
\begin{equation}
  \mathscr{L}(r;\alpha) \doteq 	-\frac{2m''G_{N}^{(4)}}{r}\,,
\end{equation}

\noindent
and the first one gives
	
\begin{equation}\label{ar}
a(r;\alpha) \doteq  \frac{1}{2\mathcal{Q}}(m' r-3m)+ c\,,
\end{equation}

\noindent
where $c$ is an integration constant.
	
We are going to consider the following form of the mass function
	
\begin{equation}\label{massm}
m(r; \alpha) \equiv M-f(r;\alpha)\,,
\end{equation}

\noindent
where $M$ is, and will always be, the ADM mass properly choosing the
integration constant in $f$. To get a simpler expression for the gauge
potential, we choose the integration constant in Eq.~\eqref{ar} to be
$c = \frac{3M}{2\mathcal{Q}}$, hence we get
	
\begin{align}
  \label{lambda}
  \lambda(r;\alpha)&=1-\frac{2MG_{N}^{(4)}}{r}+\frac{2 f(r;\alpha)G_{N}^{(4)}}{r}\,,\\
  \nonumber\\
  \label{lambdaLA}
  \mathscr{L}(r;\alpha)&\doteq \frac{2f''(r;\alpha)G_{N}^{(4)}}{r}\,,\\
  \nonumber\\
  A(r;\alpha)&\doteq 
               \frac{1}{2\mathcal{Q}}\bigg(-f'(r;\alpha)r+3f(r;\alpha)\bigg)
               \dd t.
\end{align}

In the following we are going to write the Smarr formula Eq.~\eqref{smarrrrrr}
in terms of the variables introduced in this section. This will simplify
verification across different solutions, since we only need to change the
function $f(r;\alpha)$ and the corresponding on-shell Lagrangian. We are going
to assume the following asymptotic behaviour for the function $f(r;\alpha)$
and its first derivative

\begin{equation}\label{ff'}
  f(r;\alpha)\stackrel{r\rightarrow\infty}{\rightarrow}{0},
  \qquad \qquad	f'(r;\alpha)\stackrel{r\rightarrow\infty}{\rightarrow}{0}\,.
\end{equation}

The first condition ensures that $M$ remains the ADM mass.  For the solutions
we are going to study these conditions will be enough, even if not strictly
necessary.

The temperature, the entropy and the difference in the electric potential are
given by

\begin{align}
	\label{T}
	T &= \left.\frac{\partial_{r}\lambda}{4\pi}\right|_{r_{+}}
	= \left.\frac{G_{N}^{(4)}}{4\pi}\Big(\frac{2M}{r^{2}}-\frac{2f}{r^{2}}+\frac{2f'}{r}\Big)\right|_{r_{+}},\\\nonumber\\
	\label{S}
	S &= \frac{A_{\mathcal{BH}}}{4G_{N}^{(4)}} = \frac{\pi r_{+}^{2}}{G_{N}^{(4)}}\,,
	\\\nonumber\\
	\label{phiaq}
	\Phi(r_{+})-	\Phi(\infty)&= \frac{1}{2\mathcal{Q}}(-f'r+3f)\big|_{r_{+}}\,.
\end{align}

To find $P_{k}^{H}$, we can use its definition

\begin{equation}
0=\imath_{k} H + \dd P^{H}_{k}\doteq -\imath_{k} \mathbf{L}_{\alpha} + \dd P^{H}_{k}\,,
\end{equation}

\noindent
and, since we assumed spherical symmetry, we get

\begin{equation}
  P^{H}_{k\;\theta\varphi}(r)
  \doteq
  \sin\theta \bigg(\int_{0}^r r^{2} \mathscr{L}_{\alpha} \dd r+c\bigg)\,,
\end{equation}

\noindent
where $c$ is an integration constant.  It follows that

\begin{equation}\label{lambdar}
\Lambda(r)\equiv\frac{1}{16\pi G_{N}^{(4)}}	\int_{S^{2}}   {P}^{H}_{k}
	\doteq    \frac{1}{4 G_{N}^{(4)}} \bigg(\int_{0}^r r^{2} \mathscr{L}_{\alpha} \dd r+c\bigg)\,,
\end{equation}

\noindent
and therefore

\begin{equation}\label{gammaalpha}
  \gamma [	\Lambda(r_{+})-	\Lambda(\infty)]\alpha\doteq
  -\frac{\gamma\alpha}{4G_{N}^{(4)}} \int_{r_{+}}^{\infty} r^{2} \mathscr{L}_{\alpha} \dd r\equiv I(\infty).
\end{equation}

Substituting all these ingredients in the Smarr relation Eq.~\eqref{smarrrrrr}
we get the equivalent identity

\begin{equation}\label{smarrrr}
	-\frac{\gamma\alpha}{4G_{N}^{(4)}} \int_{r_{+}}^{\infty} r^{2} \mathscr{L}_{\alpha} \dd r\doteq-\tfrac{1}{2} \bigg(f(r_{+})+f'(r_{+})r_{+}\bigg)\,.
\end{equation}

To verify the Smarr formula it is enough to compute the on-shell Lagrangian
together with the function \(f(r;\alpha)\), and to check that
Eq.~\eqref{smarrrr} is satisfied.

\subsubsection{The Einstein-Born-Infeld black hole }	

The Einstein-Born-Infeld Lagrangian \cite{Born:1934gh} for the purely electric
case ($Y=0$) can be written as
	
\begin{equation}
		\mathscr{L}= \alpha \bigg(1-\sqrt{1-\frac{2}{\alpha}X}\bigg)\,,
\end{equation}

\noindent
where, for the ansatz \eqref{A}
	
\begin{equation}\label{Xa}
  X \doteq \frac{(a')^{2}}{2}.
\end{equation}

To check Eq.~\eqref{smarrrr} we need to compute $\mathscr{L}_{\alpha}$
on-shell. As we will show in the next lines, $\mathscr{L}_{\alpha}$ can be
written in terms of $\mathscr{L}_{X}$. Hence, the best way to proceed is to
express $\mathscr{L}_{X}$ in terms of $r$. For simplicity, we introduce the
variable
	
\begin{equation}
  k(r)\equiv \frac{4\mathcal{Q}G_{N}^{(4)}}{r^{2}}\,.
\end{equation}

Using Eq.~\eqref{LXQ}, we can express $\mathscr{L}_{X}$ in terms of $k(r)$ as
	
\begin{equation}\label{LX}
  \mathscr{L}_{X}
  =
  \frac{1}{\sqrt{1-\frac{2}{\alpha}X}}
  \doteq
  \frac{1}{\sqrt{1-\frac{(a')^{2}}{\alpha}}}\doteq  -\frac{k(r)}{a'}.
\end{equation}

From the last equality we get the relation
	
\begin{equation}\label{a'k}
  (a')^{2} \doteq\frac{\alpha k^{2}}{\alpha + k^{2}}\,,
\end{equation}

\noindent
that can be exploited to express $\mathscr{L}_{X}$ in terms of $r$
	 
\begin{align}\label{lxr}
  \mathscr{L}_{X}
  \doteq
  \sqrt{\frac{\alpha + k^{2}}{\alpha}}= \frac{U(r)}{\sqrt{\alpha }r^{2}}\,,
\end{align}

\noindent
where, for simplicity,  we introduced 
	
\begin{equation}
  \beta \equiv 16 \mathcal{Q}^{2} G_{N}^{(4)2},\qquad
  U(r) \equiv \sqrt{\alpha r^{4} + \beta}\,.
\end{equation}
	
Using the first equality of Eq.~\eqref{LX} we can write the off-shell identity
	
\begin{equation}
  \frac{X}{\alpha}
  =
  \tfrac{1}{2} \bigg(1-\frac{1}{\mathscr{L}_{X}^{2}}\bigg)\,,
\end{equation}

\noindent
that can be used\footnote{The correct procedure is to compute these
  derivatives off-shell and only afterward restrict to the on-shell
  configuration. When the derivatives are taken off-shell, $X$ and $Y$ do not
  depend on $\alpha$. By contrast, differentiating after imposing the on-shell
  relations generally induces an $\alpha$-dependence in $X$ and $Y$, and gives
  a wrong result.} to write $ \mathscr{L}_{\alpha}$ in terms of
$ \mathscr{L}_{X}$
	
\begin{equation}
  \label{lalpha}
  \mathscr{L}_{\alpha}
  =
  1-  \tfrac{1}{2} \bigg(\frac{1}{\mathscr{L}_{X}}+{\mathscr{L}_{X}}\bigg).
\end{equation}

We found $\mathscr{L}_{X}$ in terms of $r$ in Eq.~\eqref{lxr}, so we can write
$\mathscr{L}_{\alpha}$ as a function of $r$.
	
Before computing the integral $I(\infty)$, we need to find explicitly the
function $f(r;\alpha)$.  The on-shell Lagrangian can be rewritten as
	
\begin{equation}
  \mathscr{L}
  \doteq
  \alpha 
  \bigg(1-\sqrt{\frac{\alpha}{\alpha + \frac{\beta}{r^{4}}}}\bigg)
  \doteq
  \frac{2 G_{N}^{(4)}f''}{r}\,,
\end{equation}

\noindent
where we used Eq.~\eqref{lambdaLA}. The last equality gives 
	
\begin{equation}
  f''(r;\alpha)
  \doteq
  \frac{\alpha r}{2 G_{N}^{(4)}}\bigg(1- \sqrt{\frac{\alpha}{\alpha + \frac{\beta}{r^{4}}}}\bigg)\,,
\end{equation}

\noindent
from which 

\begin{equation}
  \label{f'}
f'(r;\alpha) \doteq \frac{\alpha r^{2}}{4G_{N}^{(4)}}- \frac{\sqrt{\alpha}}{4 G_{N}^{(4)}}\sqrt{\alpha r^{4} + \beta }+ c_{1}.
\end{equation}

\noindent
where $c_{1}$ is an integration constant. Notice that as $r \rightarrow\infty$
	
\begin{equation}
  \label{f'inft}
f'(r;\alpha)  \stackrel{\infty}{\sim}
-\frac{\beta}{8G_{N}^{(4)}r^{2}}+....+ c_{1}
\quad
\stackrel{\infty}{\longrightarrow}
\quad
c_{1}\,,
\end{equation}

\noindent
and we choose $c_{1}=0$, consistently with Eq.~\eqref{ff'}. The function
$f(r;\alpha)$ is given by
	
\begin{equation}
  f(r;\alpha) \doteq \frac{\alpha }{12G_{N}^{(4)}}{r^{3}}
  -\frac{\sqrt{\alpha}}{4G_{N}^{(4)}}\int \sqrt{ \alpha r^{4} + {\beta}}
	\dd r+ c_{2}.
\end{equation} 

\noindent
where $c_{2}$ is another integration constant. For $r \rightarrow\infty$, the
first two terms converge to a finite value, that can be absorbed in the
integration constant, so $f\stackrel{r \rightarrow\infty}{\longrightarrow}0$,
consistently with Eq.~\eqref{ff'}.
	
Now we can compute the integral $I(\infty)$ defined in Eq.~\eqref{gammaalpha},
considering $\gamma =-2$ for the Einstein-Born-Infeld theory, writing
$\mathscr{L}_{\alpha}$ as a function of $r$ from Eqs.~\eqref{lxr} and
\eqref{lalpha}, and introducing a finite radius $R>r_{+}$ which is then sent to
infinity $R\rightarrow\infty$:
	
\begin{equation}
  \label{IR}
  I(R)
  = 
  \frac{\alpha}{2G_{N}^{(4)}} 
  \int_{r_{+}}^{R} r^{2} \dd r
  -	\frac{\sqrt{\alpha}}{4G_{N}^{(4)}} 
  \int_{r_{+}}^{R} U(r) \dd r
  -	\frac{\alpha^{3/2}}{4G_{N}^{(4)}} 
  \int_{r_{+}}^{R} \frac{r^{4} }{U(r)}\dd r.
\end{equation}

It is easy to check that the function $U(r)$ satisfies the relation
	
\begin{equation}
	 \frac{r^{4}}{U(r)}= \frac{r U'(r)}{2\alpha}\,,
\end{equation}

\noindent
that can be used to rewrite the third integral  in Eq.~\eqref{IR} as
	
\begin{equation}
  \int_{r_{+}}^{R} \frac{r^{4} }{U(r)}\dd r
  =
  \int_{r_{+}}^{R} \frac{r U'(r)}{2\alpha}\dd r
  =
  \frac{1}{2\alpha}(RU(R)-r_{+}U(r_{+}))
  -\frac{1}{2\alpha}\int_{r_{+}}^{R}U(r)\dd r\,.
\end{equation}

Eq.~\eqref{f'} can be rearranged as
	
\begin{equation}\label{alpU}
  \sqrt{\alpha} U(r)=\alpha r^{2} -4 G_{N}^{(4)} f'\,,
\end{equation}

\noindent
to rewrite the second integral of Eq.~\eqref{IR} as
	
\begin{equation}
  \frac{\sqrt{\alpha}}{8G_{N}^{(4)}} 
  \int_{r_{+}}^{R} U(r) \dd r
  = 	\frac{{\alpha}}{8G_{N}^{(4)}} \int_{r_{+}}^{R} r^{2} \dd r
  -\tfrac{1}{2}(f(R)-f(r_{+}))\,.
\end{equation}

Combining these results and using Eq.~\eqref{f'}, we get the final expression
	
\begin{equation}
  I(R)
  \doteq
  {\frac{R}{2}f'(R)}
  -
  \frac{{\alpha}}{8G_{N}^{(4)}} r_{+}^{3}
  +\tfrac{1}{2} f(R)-\tfrac{1}{2} f(r_{+})
  +\frac{\sqrt{\alpha}}{8G_{N}^{(4)}} r_{+} U(r_{+})\,.
\end{equation}
	
Now we can take the limit $R\rightarrow\infty$:
$f(R)\stackrel{R \rightarrow\infty}{\longrightarrow}0$ as we already
discussed, using Eq.~\eqref{f'inft} we have
$ \frac{R}{2}f'(R)\stackrel{R \rightarrow\infty}{\longrightarrow}0$, and using
Eq.~\eqref{alpU} we get
	
\begin{equation}
I(\infty)\doteq -\tfrac{1}{2} \bigg(f(r_{+})+r_{+} f'(r_{+})\bigg)\,,
\end{equation}

\noindent
which is precisely the  relation \eqref{smarrrr}.

\subsubsection{The Einstein-ModMax black hole}

The ModMax Lagrangian is \cite{Bandos:2020jsw, Kosyakov:2020wxv}

\begin{equation}
  \mathscr{L}
  =
  \cosh \alpha X +\sinh \alpha\sqrt{X^{2}+Y^{2}}.
\end{equation}

In this theory $\alpha$ is dimensionless. Hence, it does not rescale and
$ \gamma=0$. This is a trivial example where the Smarr formula is not modified
by the presence of the coupling constant. In this section we are going to
consider the electric case and in section \ref{magneticmodmax} we will explore
the magnetic one.

For the purely electric solution $Y\doteq0$, so

\begin{equation}
  \mathscr{L}\doteq e^{\alpha} X\,,
\end{equation}
      
\noindent
and, using Eq.~\eqref{LXQ},

\begin{equation}
  \mathscr{L}_{X}
  = \cosh \alpha +\sinh \alpha\frac{X}{\sqrt{X^{2}+Y^{2}}}
  \doteq 	e^{\alpha} \doteq-\frac{k(r)}{a'}\,,
\end{equation}
      
\noindent
from which we get the relation

\begin{equation}\label{a'}
  a'\doteq-e^{-\alpha}k(r).
\end{equation}
      
Using Eqs.~\eqref{Xa} and \eqref{a'} we can express $ \mathscr{L}_{\alpha}$ as
a function of $r$

\begin{equation}
  \mathscr{L}_{\alpha}
  =
  \sinh \alpha X +\cosh \alpha\sqrt{X^{2}+Y^{2}}
  \doteq e^{\alpha} X \doteq \frac{e^{-\alpha}}{2}k^{2}.
\end{equation}

From Eq.~\eqref{lambdaLA} we have

\begin{equation}
  f'' \doteq \frac{4 \mathcal{Q}^{2}G^{(4)}_{N}e^{-\alpha}}{r^{3}}\,,
\end{equation}

\noindent
and, integrating,

\begin{equation}
f'\doteq -\frac{2 \mathcal{Q}^{2}G^{(4)}_{N}e^{-\alpha}}{r^{2}}+c_{1}\,,
\end{equation}

\noindent
where $c_{1}$ is an integration constant that we set to zero in order to have
$f'\stackrel{r\rightarrow \infty}{\rightarrow}0$. Integrating again we get

\begin{equation}
  f \doteq\frac{2 \mathcal{Q}^{2}G^{(4)}_{N}e^{-\alpha}}{r}+c_{2} \,,
\end{equation}

\noindent
and we choose $c_{2}=0$ to get
$f\stackrel{r\rightarrow \infty}{\rightarrow}0$.  Using these relations it is
easy to see that

\begin{equation}
	-\tfrac{1}{2} ( f+r f')\big|_{r_{+}}\doteq0\,,
\end{equation}

\noindent
as expected by the Smarr equation \eqref{smarrrr} with $\gamma=0$. Notice that
a problem may arise if $\int_{r_{+}}^{\infty} r^{2} \mathscr{L}_{\alpha} \dd r$
diverges; however, one easily verifies that it converges.

\subsection{Magnetic black holes}\label{magnetic}

\subsubsection{The ansatz}

Consider the following metric
	
\begin{equation}
  \dd s^{2} = \lambda(r) \dd t^{2} - \frac{1}{\lambda(r) }
  \dd r^{2} -r^{2} (\dd \theta^{2} +\sin ^{2}\theta \dd \varphi^{2})\,,
\end{equation}

\noindent
with gauge potential
	
\begin{equation}\label{a}
  A
  =
  A_{\varphi}(\theta)\dd \varphi
  \equiv
  -4 G^{(4)}_{N}\mathcal{P} \cos \theta \dd \varphi\,.
\end{equation}

Thus, we are considering a spherically symmetric black hole with magnetic
charge defined in Eq.~\eqref{magncharge}. As in the previous section, we are
going to assume that the Lagrangian $\mathscr{L}$ is just a function of $X$
and does not depend on $Y$.
	
The non-linear Maxwell equations \eqref{EL} are automatically satisfied using
the ansatz in Eq.~\eqref{a}. Using again Eq.~\eqref{matrix}, the Einstein
equations give rise to two independent equations
	
\begin{align}
  \label{einsteinNLMagnetic}
  \frac{\lambda'}{r}+\frac{\lambda-1}{r^{2}}-\frac{\mathscr{L}}{2}&\doteq 0\,,\\
  \nonumber\\\label{einsteinNLMagnetic2}
  \lambda'' + \frac{2\lambda '}{r} -\mathscr{L}-16(G^{(4)}_{N})^{2}\mathcal{P}^{2}\frac{\mathscr{L}_{X}}{r^{4}}&\doteq 0\,.
\end{align}

Eq.~\eqref{einsteinNLMagnetic} gives $\mathscr{L}$ on-shell as a function of
$r$
	
\begin{equation}
  \mathscr{L}
  \doteq
  2 \bigg(\frac{\lambda'}{r}+\frac{\lambda-1}{r^{2}}\bigg)\,,
\end{equation}

\noindent
and, using the  on-shell Lorentz scalar 
	
\begin{equation}\label{X}
  X \doteq - \frac{8 (G^{(4)}_{N})^{2}\mathcal{P}^{2}}{r^{4}}\,,
\end{equation}

\noindent
we can compute $\mathscr{L}_{X} = {\mathscr{L}'}/{X'}$ on-shell and
substituting it into Eq.~\eqref{einsteinNLMagnetic2} we see that the latter is
automatically satisfied. In terms of the mass function \eqref{massm} the
on-shell Lagrangian is
	
\begin{align}
\label{lf'}
  \mathscr{L}
  \doteq
  \frac{4f' G_{N}^{(4)}}{r^{2}}\,.
\end{align}

As we did for the electric case, we are going to write the Smarr formula
Eq.~\eqref{smarrrrrr} in terms of $f(r;\alpha)$ and the on-shell Lagrangian in
order to simplify the verification across different solutions.  To find
$\tilde{P}_{k} $ we can use the dual momentum map equation
	
\begin{equation}
\label{tildeP}
\imath_{k} \star \Pi + \dd \tilde{P}_{k}
\doteq
0\,,
\qquad \Rightarrow \qquad
\tilde{P}_{k}(r)
\doteq
\tilde{P}_{k}(0)
+\int_{0}^{r}  \frac{4G^{(4)}_{N}\mathcal{P}}{ r^{2}}\mathscr{L}_{X} \dd r.
\end{equation}

Using Eqs.~\eqref{T}, \eqref{S}, \eqref{gammaalpha}, the Smarr formula
\eqref{smarrrrrr} is equivalent to the following equation
	
\begin{equation}
  \label{magneticsmarr}
		{-\frac{\gamma\alpha}{4G_{N}^{(4)}} \int_{r_{+}}^{\infty} r^{2} \mathscr{L}_{\alpha} \dd r+4 G^{(4)}_{N} \mathcal{P}^{2} \int_{r_{+}}^{\infty} \frac{1}{r^{2}}\mathscr{L}_{X}\dd r\doteq f(r_{+})-f'(r_{+})r_{+}}\,.
\end{equation}

\subsubsection{The Einstein-Hayward black hole}

The following Lagrangian belongs to the Hayward class defined in
\cite{Fan:2016hvf} in terms of a parameter $\mu $ that we set to be $\mu = 1$
throughout this work:

\begin{equation}\label{HBH}
\mathscr{L} = \frac{-X}{[1+(-\alpha X)^{\frac{1}{4}}]^{2}}\,,
\end{equation}

\noindent
where $\gamma =2$.  Using Eq.~\eqref{X}, we can write it on-shell as

\begin{equation}
  \mathscr{L}\doteq \frac{\sigma^{4}}{  [r^{2}+\alpha^{\frac{1}{4}}\sigma r]^{2}}\,,
\end{equation}

\noindent
where we defined for simplicity

\begin{equation}\label{sigma}
\sigma \equiv \bigg[8 (G^{(4)}_{N})^{2}{\mathcal{P}^{2}}\bigg]^{\frac{1}{4}}\,.
\end{equation}

For this ansatz

\begin{align}
  \mathscr{L}_{\alpha}
  & =
    -\frac{(-X)^{\frac54}}{2\alpha^{\frac34}[1+(-\alpha X)^{\frac{1}{4}}]^{3}}
    \doteq
    -\frac{\sigma^5r}{2\alpha^{\frac34} [r^{2}+\alpha^{\frac{1}{4}}\sigma r]^{3}}\,,
  \\
                        & \nonumber \\
  \mathscr{L}_{X}
  & =
    -\frac{2+(-\alpha X)^{\frac{1}{4}}}{2[1+(-\alpha X)^{\frac{1}{4}}]^{3}}
    \doteq
    -\frac{2r^{3}+\alpha^{\frac{1}{4}}{\sigma}{r}^{2}}{2 [r+\alpha^{\frac{1}{4}}{\sigma}]^{3}}\,.
\end{align}

Integrating Eq.~\eqref{lf'} we get

\begin{equation}
  f
  \doteq
  -\frac{\sigma^{4}}{4 G^{(4)}_{N}}
  \frac{1}{(r+\alpha^{\frac{1}{4}}{\sigma})}+c\,,
\end{equation} 

\noindent
where, as usual, we choose $c=0$ so
$f \stackrel{r \rightarrow \infty}{\rightarrow}0$.

Using Eq.~\eqref{gammaalpha}, we get

\begin{align*}
  \gamma [	\Lambda(r_{+})-	\Lambda(\infty)]\alpha\doteq 
  \frac{1}{G_{N}^{(4)}}\frac{\alpha^{\frac{1}{4}}\sigma^5}{8(r_{+}+\alpha^{\frac{1}{4}}\sigma)^{2}}\,,
\end{align*}

\noindent
and the second integral in Eq.~\eqref{magneticsmarr}  gives

\begin{equation}
	 \int_{r_{+}}^{\infty} \frac{1}{r^{2}}\mathscr{L}_{X}\dd r= -\frac{4r_{+}+3 \alpha^{\frac{1}{4}}\sigma}{4(r_{+}+\alpha^{\frac{1}{4}}\sigma)^{2}}\,.
\end{equation}

Combining these results, it is easy to check that Eq.~\eqref{magneticsmarr} is
satisfied, and therefore we verified the Smarr formula \eqref{smarrrrrr}.

\subsubsection{The Einstein-ModMax black hole\label{magneticmodmax}}

For the purely magnetic solution  with $Y\doteq0$, the on-shell Lagrangian is 

\begin{equation}
  \mathscr{L}
  \doteq
  e^{\alpha} X\doteq -\frac{e^{\alpha}\sigma^{4}}{r^{4}}
\end{equation}
      
\noindent
where we used Eq.~\eqref{X}, and, as in the electric case, $\mathscr{L}_{X}$
is given by

\begin{equation}
  \mathscr{L}_{X}\doteq 	e^{\alpha} \,.
\end{equation}

From Eq.~\eqref{lf'} we get 

\begin{equation}
f' \doteq -\frac{e^{\alpha}\sigma^{4}}{4G^{(4)}_{N}r^{2}}\,,
\end{equation}

\noindent
that satisfies $f'\stackrel{r\rightarrow\infty}{\rightarrow}{0}$, 
and integrating we get

\begin{equation}
  f \doteq \frac{e^{\alpha}\sigma^{4}}{4G^{(4)}_{N}r}+c\,,
\end{equation}

\noindent
where we set $c=0$ so $f\stackrel{r\rightarrow\infty}{\rightarrow}{0}$. The
second integral in Eq.~\eqref{magneticsmarr} gives

\begin{equation}
  \int_{r_{+}}^{\infty} \frac{1}{r^{2}}\mathscr{L}_{X}\dd r
  =
  \frac{e^{\alpha}}{r_{+}}\,.
\end{equation}

Combining these results with $\gamma=0$, it is easy to check that
Eq.~\eqref{magneticsmarr} is satisfied, and therefore we verified the Smarr
formula \eqref{smarrrrrr}.

\subsubsection{Regular black holes: the Einstein-Bardeen solution}\label{bardeen}

Consider the following Lagrangian \cite{Ayon-Beato:2000mjt, Fan:2016hvf}

\begin{equation}
	\mathscr{L} =\frac{1}{\alpha} \bigg[\frac{\sqrt{-\alpha X}}{[1+\sqrt{-\alpha X}]}\bigg]^{\frac52}.
\end{equation}

\noindent
with $\gamma=2$. Using Eq.~\eqref{X} we have

\begin{align}
  \mathscr{L}&\doteq \frac{{\alpha^{\frac{1}{4}} }\sigma^5}{[r^{2}+\sqrt{\alpha }\sigma^{2}]^{\frac52}}\,,
  \\\nonumber\\
  \mathscr{L}_{\alpha}  &= -
                          \frac{(-\alpha X)^{\frac54}[4 \sqrt{-\alpha X}-1]}{4\alpha^{2}[1+\sqrt{-\alpha X}]^{\frac72}}\doteq \frac{\sigma^5}{4 \alpha^{\frac34}}
                          \frac{r^{2}-4 \sqrt{\alpha}\sigma^{2}}
                          { [r^{2}+ \sqrt{\alpha}\sigma^{2}]^{\frac72}}\,,
  \\\nonumber\\
  \mathscr{L}_{X} & = -\frac54  \frac{(-\alpha X)^{\frac{1}{4}}}{[\sqrt{-\alpha X }+1]^{\frac72}}
                    \doteq -\frac{\sigma \alpha^{\frac{1}{4}}5}{4}\frac{r^6}{[r^{2}+ \sqrt{\alpha}\sigma^{2}]^{\frac72}}\,.
\end{align}

We can find $f$ integrating Eq.~\eqref{lf'}

\begin{equation}\label{f}
  f(r;\alpha)\doteq
  \frac{\sigma^{3}}{12G^{(4)}_{N}\alpha^{\frac{1}{4}}} \frac{r^{3}}{[r^{2}+ \sqrt{\alpha}\sigma^{2}]^{\frac{3}{2}}}+c.
\end{equation} 

Setting the integration constant
$c= - \frac{\sigma^{3}}{12G^{(4)}_{N}\alpha^{\frac{1}{4}}} $ ensures
$ f(r;\alpha)\stackrel{r\rightarrow\infty}{\rightarrow}{0}$. Substituting this
expression in $\lambda$, we get
 
\begin{equation}\label{lambdaralpha}
  \lambda(r;\alpha)
  =
  1 -\frac{2(M-c) G^{(4)}_{N}}{r}
  +\frac{\sigma^{3} r^{2}}{6\alpha^{\frac{1}{4}}[r^{2}+ \sqrt{\alpha}\sigma^{2}]^{\frac{3}{2}}}\,,
\end{equation}
      
\noindent
from which it is clear that regular configurations are obtained when the ADM
mass satisfies

\begin{equation}\label{adm}
  M= -\frac{\sigma^{3}}{12G_{N}^{(4)} \alpha^{\frac{1}{4}}}.
\end{equation}

Note that, to ensure the mass is real and positive, the parameter $\sigma$
given by the fourth root in Eq.~\eqref{sigma} must be negative.

Using  Eq.~\eqref{gammaalpha}, we get

\begin{equation}
  \gamma [	\Lambda(r_{+})-	\Lambda(\infty)]\alpha
  \doteq 
   \frac{M}{2}\bigg(\frac{r_{+}^{3}[r_{+}^{2} +4
     \sqrt{\alpha}\sigma^{2}]}{[r_{+}^{2}
     +\sqrt{\alpha}\sigma^{2}]^{\frac{5}{2}}}-1\bigg)\,,
\end{equation}

\noindent
and the second integral in Eq.~\eqref{magneticsmarr}  gives

\begin{equation}
  \int_{r_{+}}^{\infty}\frac{1}{ r^{2}}\mathscr{L}_{X} \dd r
  \doteq
  \frac{1}{4\sigma \alpha^{\frac{1}{4}}}
  \bigg(\frac{r_{+}^5}{[{r_{+}^{2}+\sqrt{\alpha}\sigma^{2}}]^{\frac{5}{2}}}
  -1\bigg)\,.
\end{equation}
      
Combining these results it is easy to check that Eq.~\eqref{magneticsmarr} is
satisfied, and therefore we verified the Smarr formula \eqref{smarrrrrr}.

As we have explained in the Introduction, the reason for considering
regular configurations in this section is that the generalized Komar
charge allows us to study when and how there is no singularity. We
will not repeat here the arguments presented in the Introduction and
we will just proceed to examine from that point of view the absence of
singularity in the regular Bardeen solution comparing it with the
singular, magnetically charged Reissner--Nordstr\"om black hole
studied in Appendix~\ref{Appendix}. In particular, we will compute
each contribution to the Komar integral and analyze their behaviour in
the limits $r\rightarrow\infty$ and $r\rightarrow 0$.

It is convenient to write the generalized Komar charge as in
Eq.~(\ref{k0kned}) and the equation that expresses its conservation in
the form of Eq.~(\ref{eq:nonconservationofK0}).
In particular,
if we choose a spatial volume $\Sigma^{3}_{r}$  with boundary
$\partial \Sigma^{3}_{r}=S^{2}_{r}\cup S^{2}_\infty$, where $S^{2}_{r}$
is a 2-sphere of radius $r$, integrating $\dd \mathbf{K}[k]\doteq 0$,
with $\mathbf{K}[k]$ given by Eq.~\eqref{k0kned}, over $\Sigma^{3}_{r}$
and applying Stokes' theorem, we get

\begin{equation}\label{kkk}
  \int_{S^{2}_{\infty}}\mathbf{K}_{0}[k]
  \doteq
  \int_{S^{2}_{r}} \mathbf{K}_{0}[k]
  +\int_{\Sigma^{3}_{r}} \mathbf{Z}[k]\,,
\end{equation}

\noindent
where

\begin{equation}
  \int_{S^{2}_{r}} \mathbf{K}_{0}[k]
  =
  \frac{r^{2} \lambda'}{4 G_{N}^{(4)}}\,.
\end{equation}

\noindent
The integral over $S^{2}_{\infty}$ always gives $M/2$ for
asymptotically-flat solutions. Therefore Eq.~\eqref{kkk} becomes

\begin{equation}\label{kkkk}
  \frac{M}{2}
  \doteq
  \frac{r^{2} \lambda'}{4 G_{N}^{(4)}}
  +\int_{\Sigma^{3}_{r}} \mathbf{Z}[k]\,.
\end{equation}

$\mathbf{Z}[k]=0$ for the Schwarzschild solution. Then, in order to
get the result in the left-hand side, $\lambda' \sim 1/r^{2}$ and
$\lambda$ must be singular as $r \to 0$. A similar behavior occurs for
the Reissner--Nordstr\"om black hole (see Appendix~\ref{Appendix}):
$\mathbf{Z}[k]\neq 0$ and we get $\lambda' \sim 1/r^{2}-{1}/{r^{3}}$
which means that $\lambda$ is again singular as $r \to 0$. As we will
see in the next paragraph, in the Bardeen solution considered in
Eq.~\eqref{lambdaralpha}, instead, the term $r^{2} \lambda'$ in
Eq.~\eqref{kkkk} will go to zero as $r \to 0$, so $\lambda$ will be
regular and the integral of $\mathbf{Z}[k]$ in Eq.~\eqref{kkkk} will
be interpreted as a source of the ADM mass in the limit $r \to 0$.

\paragraph{Generalized Komar integral}

For the Einstein--Bardeen theory considered in this section, the Komar charge
is given by Eq.~\eqref{komar} without the electric term
	
\begin{equation}\label{komarbardeen}
  \mathbf{K}[k]=
  \frac{1}{16\pi G_{N}^{(4)}}
  \bigg\{-\star (e^{a} \wedge e^{b})P_{k ab}
  -	\tfrac{1}{2}\tilde{P}_{k}  F
  +\frac{\gamma}{2}\alpha P_{k}^{H}
  \bigg\}.
\end{equation}

We proceed to calculate the contribution of each term by integrating over a
sphere of radius $r$.  For the Killing vector $k = \partial_{t}$, the first
term gives
	
\begin{equation}
  -\star (e^{a} \wedge e^{b})P_{k ab}
  =
  -\star \dd \mathbf{k}= \sqrt{|g|}\lambda' \dd\theta \wedge \dd \varphi\,,
\end{equation}

\noindent
so we get 
	 
\begin{equation}\label{eep}
  \mathbf{K}_{0}[k]
  =
  \frac{1}{16\pi G_{N}^{(4)}}
  \int_{S^{2}_{r}}\big\{-\star (e^{a} \wedge e^{b})P_{k ab}\big\}
  = 
\frac{M}{2}\frac{r^{3}[r^{2}-2 \sqrt{\alpha}\sigma^{2}]}{[r^{2}+\sqrt{\alpha}\sigma^{2}]^{\frac{5}{2}}}\,.
\end{equation}

Note that in the limit $r\rightarrow\infty$ the gravitational term
\eqref{eep} approaches $M/2$, as expected.  In the limit
$r\rightarrow 0$, however, the two cases differ (see
Table~\ref{bardeenRN}): for the Bardeen solution, Eq.~\eqref{eep}
gives zero; the corresponding term in the Reissner--Nordstr\"om case
gives rise to a divergent term that is canceled by the magnetic
contribution to the Komar integral.
	
Using Eq.~\eqref{tildeP} and Eq.~\eqref{lambdar} we get the
dual-momentum map as a function of $r$ and $ \Lambda (r)$
	
\begin{align}
  \tilde{P}_{k}(r)
  & \doteq
    \frac{
    G_{N}^{(4)}\mathcal{P}}{\alpha^{\frac{1}{4}}\sigma}\bigg(1-\frac{r^5}{(r^{2}+\sqrt{\alpha}\sigma^{2})^{\frac{5}{2}}}\bigg)\,,
  \\
& \nonumber \\
\label{lambdabardeen}
  \Lambda (r)
  & =
    \frac{M}{4\alpha}\bigg(\frac{r^{3}[r^{2} +4 \sqrt{\alpha}\sigma^{2}]}{[r^{2} + \sqrt{\alpha}\sigma^{2}]^{\frac{5}{2}}}-1\bigg)\,,
\end{align}
        
\noindent
where we set
$ \tilde{P}_{k}(0) =\frac{
  G_{N}^{(4)}\mathcal{P}}{\alpha^{\frac{1}{4}}\sigma}$ to get
$ \tilde{P}_{k}(r) \stackrel{r\rightarrow\infty}{\rightarrow}0$ and
$\Lambda(0)= -\frac{M }{4\alpha}$ to ensure
$\Lambda(r)\stackrel{r\rightarrow\infty}{\rightarrow}0$. With this choice, the
second and the third terms in Eq.~\eqref{komarbardeen} give
	
\begin{align}\label{tildePF}
  \frac{1}{16\pi G_{N}^{(4)}}
  \int_{S^{2}_{r}}\bigg\{
  -	\tfrac{1}{2}\tilde{P}_{k}(r)  F\bigg\} = 
  \frac{3M}{4}\bigg(1-\frac{r^5}{[{r^{2}+\sqrt{\alpha}\sigma^{2}}]^{\frac{5}{2}}}\bigg)\,\\
  \nonumber\\
		\label{alphaP}
  \frac{1}{16\pi G_{N}^{(4)}}\int_{S^{2}_{r}}
  \bigg\{\frac{\gamma}{2}\alpha P_{k}^{H}
  \bigg\}
  = 
  \frac{M}{4}\bigg(\frac{r^{3}[r^{2} +4 \sqrt{\alpha}\sigma^{2}]}{[r^{2} + \sqrt{\alpha}\sigma^{2}]^{\frac{5}{2}}}-1\bigg)\,.
\end{align}

Notice that in the limit $r\rightarrow\infty$, both integrals vanish
for the Bardeen and the Reissner--Nordstr\"om solution. In the limit
$r \to 0$, for the Reissner--Nordstr\"om black hole, the magnetic
contribution to the Komar integral yields a divergent term that
cancels the gravitational one mentioned above, thereby making the
Komar integral independent of $r$, as expected from Stokes' theorem.
By contrast, in the Bardeen solution each contribution to the Komar
integral remains finite as $r\rightarrow 0$: the gravitational one
tends to zero, while the other two contributions are proportional to
$M$ with coefficients whose sum is $1/2$. This smooth behaviour is a
consequence of the nonlinearity of the theory. Indeed, for the Bardeen
solution, when $r\rightarrow0$, the surface integral in the right-hand
side of Eq.~\eqref{kkkk} vanishes and we get
	
\begin{equation}
  \label{m2K}
  \frac{M}{2}
  =
  \int_{S^{2}_{\infty}}\mathbf{K}_{0}[k]
  =
  \int_{\Sigma^{3}} \mathbf{Z}[k]\,,
\end{equation}

\noindent
from which, it is clear that, in this limit, $\mathbf{K}_{\rm NLED}[k]$ is the
only source of the ADM mass.

\begin{table}[!ht]
	\centering
	\begin{tabular}{c|cc|cc}
		\hline\hline
		Komar integral terms & \multicolumn{2}{c|}{Bardeen} & \multicolumn{2}{c}{Reissner--Nordstr\"om} \\
		& $r\rightarrow\infty$ & $r\rightarrow 0$ & $r\rightarrow\infty$ & $r\rightarrow 0$ \\
		\hline\hline
		$\int_{S^{2}_{r}}\mathbf{K}_{0}[k]$& $\dfrac{M}{2}$ & $0$ & $\dfrac{M}{2}$ & $-\infty$ \\
		\hline
		$	\frac{1}{16\pi G_{N}^{(4)}}
		\int_{S^{2}_{r}}\{
		-	\tfrac{1}{2}\tilde{P}_{k}(r)  F\}$ & $0$ & $\dfrac{3M}{4}$ & $0$ & $+\infty$ \\
		\hline
		$	\frac{1}{16\pi G_{N}^{(4)}}\int_{S^{2}_{r}}
		\{\frac{\gamma}{2}\alpha P_{k}^{H}
		\}$ & $0$ & $-\dfrac{M}{4}$ & $/$  & $/$ \\
		\hline\hline
	\end{tabular}
	\caption{Asymptotic behaviour of the Komar integral contributions in the Bardeen and Reissner--Nordstr\"om  solutions.}
	\label{bardeenRN}
\end{table}

Combining these results, the $r$-dependent terms in Eqs.~\eqref{tildePF} and
\eqref{alphaP} cancel the contribution in Eq.~\eqref{eep}, leaving

\begin{equation}
	{\begin{aligned}
			\int_{S^{2}_{r}}	\mathbf{K}(k) =
			\frac{M}{2}\,,
		\end{aligned}
	}
\end{equation}
independently of the radius, as expected from Stokes' theorem.

To conclude this section, in the next few paragraphs we are going to study the
horizons, the energy conditions, and the acausal region of the regular Bardeen
solution \cite{Man:2013hpa,Tzikas:2018cvs}.

\paragraph{Thermodynamics}

To study the horizons of this solution, we need to find the zeros of $g_{tt}$
which is given in Eq.~\eqref{lambdaralpha}

\begin{equation}\label{eq:lambda}
1-\frac{2MG_{N}^{(4)} r^{2}}{[r^{2}+\sqrt{\alpha}\,\sigma^{2}]^{3/2}}=0\,.
\end{equation}

To simplify the notation, we can introduce

\begin{equation}
  c\equiv \sqrt{\alpha}\,\sigma^{2}>0, \qquad x\equiv r^{2}\ge 0\,.
\end{equation}
      
Solving  Eq.~\eqref{eq:lambda} is equivalent to find the roots of 

\begin{equation}\label{cubic}
	 F(x)\equiv (x+c)^{3}-4M^{2}(G_{N}^{(4)})^{2} x^{2} \,.
\end{equation}
       
\begin{figure}[h]
	\centering
	\includegraphics[width=1\textwidth]{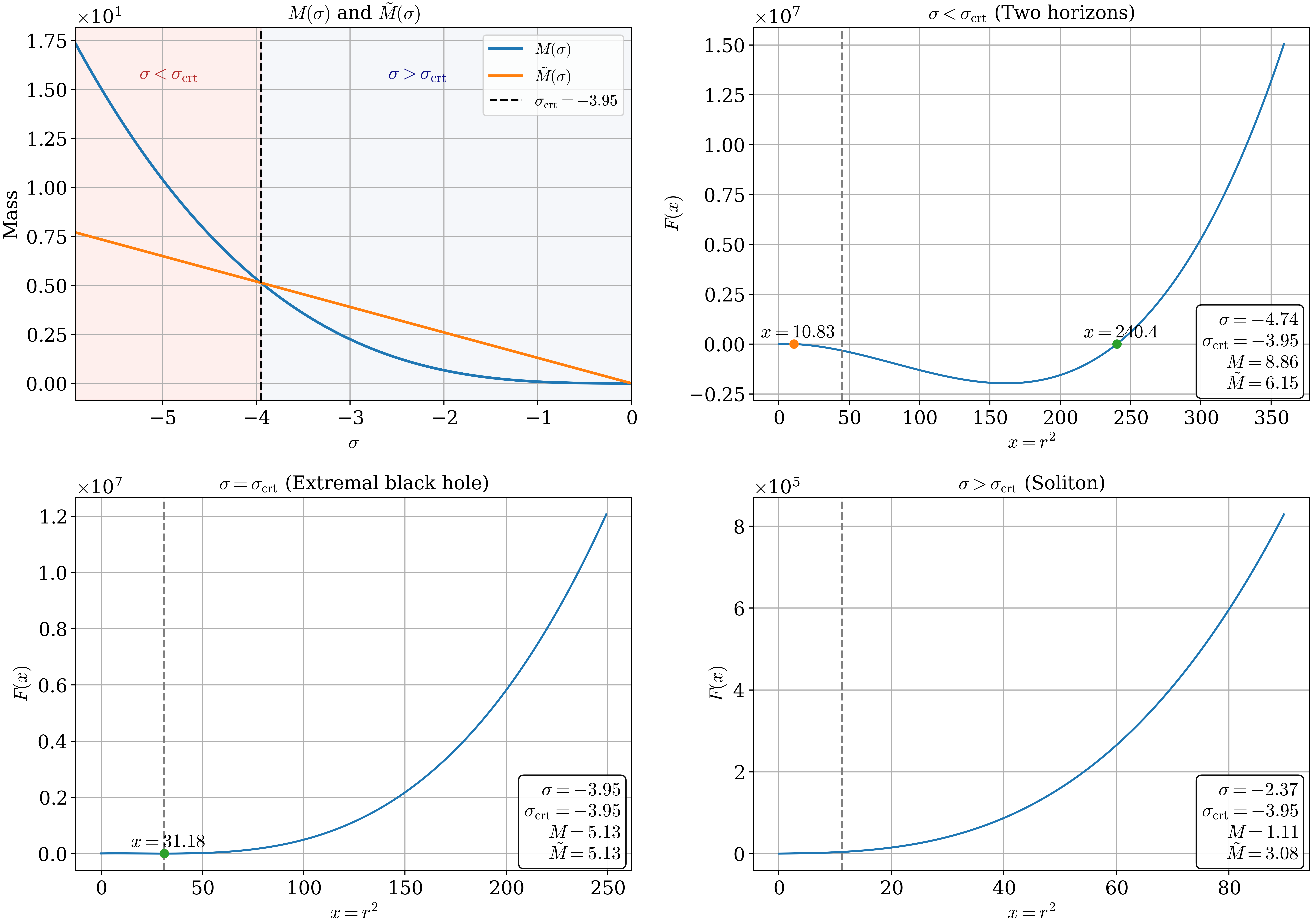} 
	\caption{Top-left: $M(\sigma)$ and $\tilde M(\sigma)$, with shaded regions indicating $\sigma\gtrless\sigma_{\rm crt}$.
		Other panels: $F(x)$ plotted for representative values of $\sigma$ showing  the two-horizon case, the extremal case, and the soliton  case. In all plots we set $\alpha=1$ and $G_{N}^{(4)}=1$. }
	\label{fig:single}
\end{figure}

This function satisfies $F(0)=c^{3}>0$ and $ \lim_{x\to\infty}F(x)=+\infty$,
hence there are only three possibilities (see Fig.~\ref{fig:single})

\begin{itemize}
	
\item $F(x)$ touches zero once and then grows up again for some value of $x$,
  $\tilde{x}$. This happens when
	
\begin{equation}\label{inter}
		F(\tilde{x})=0, \qquad F'(\tilde{x})=0,
\end{equation}

\noindent
that corresponds to $\tilde{x}$ being a double root of $F$.  When
$M=\tilde{M}$, where $\tilde{M}$ denotes the value of the ADM mass for which
condition \eqref{inter} can be satisfied, there is a single degenerate horizon
for a certain $r=\tilde{r}$. This is the extremal case ($T=0$).

\item If $F(x)>0$ for all $x\ge0$, $F(x)$ has no real roots and the spacetime
  is horizonless. This is a soliton. This happens when $M<\tilde{M}$.
	
\item If $F(x)$ dips below zero and returns to $+\infty$, $F(x)$ has two real
  and positive roots, that are the inner and outer horizons of the regular
  black hole. This happens when $M>\tilde{M}$.
	
\end{itemize} 

The condition $F'(x)=0$ gives

\begin{equation}
	M^{2}=\frac{3(x+c)^{2}}{8x(G_{N}^{(4)})^{2}}\,,
\end{equation}

\noindent
and, substituting this into $F(x)=0$ we get the extremal radius,

 \begin{equation}
   \tilde{r}
   =
   \sqrt{\tilde{x}}=\sqrt{2c}=\sqrt{2}\,\alpha^{1/4}\,|\sigma|\,,
\end{equation}

\noindent
and the corresponding mass 

\begin{equation}\label{tildeadm}
\tilde{M} = \frac{3\sqrt{3}}{4}\frac{\alpha^{1/4}|\sigma|}{G_{N}^{(4)}}\,,
\end{equation}

\noindent
as functions of $\sigma$, that we recall it is related to the magnetic charge via Eq.~\eqref{sigma}. For  $M=\tilde{M}$, $F(x)$ can be factorized as 

\begin{equation}
F(x)|_{\tilde{M}}=  \tfrac{1}{4} (x-2c)^{2}(c+4x)\,,
\end{equation}

\noindent
from which it is manifest that $\tilde{x}=2c$ has multiplicity 2.  Comparing
Eq.~\eqref{adm} with Eq.~\eqref{tildeadm}, we can derive the value of $\sigma$
for which the ADM mass is equal to $\tilde{M}$. We call it $\sigma_{\rm crt}$
and it can be written in terms of the coupling constant as
 
\begin{equation}
\sigma_{\rm crt} = -\,(3^{5}\alpha)^{\frac{1}{4}}\,.
\end{equation}

The relationship between the masses $M$ and $\tilde{M}$ fixes the range of
$\sigma$ as follows

\begin{equation}
  \displaystyle
  \begin{cases}
    M > \tilde{M}
    \quad\Rightarrow\quad
    \sigma < \sigma_{\rm crt} \\[6pt]
    M = \tilde{M}
    \quad\Rightarrow\quad
    \sigma = \sigma_{\rm crt} \\[6pt]
    M < \tilde{M}
    \quad\Rightarrow\quad
    \sigma_{\rm crt} < \sigma < 0\,.
  \end{cases}
\end{equation}

A plot of $M$ and $\tilde M$ as functions of $\sigma$ is given in
Fig.~\ref{fig:single}.  Notice that when $M< \tilde{M}$ we get a soliton
spacetime, that is a horizonless gravitational configuration. Instead, for the
Reissner--Nordstr\"om black hole, when\footnote{Using the results in
  Appendix~\ref{Appendix}, it is easy to prove that
  $\tilde{M}^{RN}= \frac{\sigma^{2}}{\sqrt{2}G_{N}^{(4)}}$.}
$M< \tilde{M}^{RN}$ we get a naked singularity.

If $r_{+}$ is the outer horizon, from Eq.~\eqref{cubic} one can solve for the
mass in terms of $r_{+}$
	
\begin{equation}\label{M_of_{r}}
  M(r_{+})
  =
  \frac{(r_{+}^{2}+c)^{\frac{3}{2}}}{2G_{N}^{(4)}r_{+}^{2}}\,,
\end{equation}

\noindent
that has a minimum at $r_{+}^{2}=\tilde{r}^{2}=2c$, giving the minimum mass
$\tilde{M}$ found above. Using Eq.~\eqref{M_of_{r}} we can compute
$\lambda'(r)$ and get the temperature as a function of $r_{+}$
	
\begin{equation}\label{t}
  T=\frac{\lambda'(r_{+})}{4\pi}=\frac{r_{+}^{2}-2c}{4\pi\,r_{+}(r_{+}^{2}+c)}.
\end{equation}

\begin{figure}[h]
	\centering
	\includegraphics[width=1\textwidth]{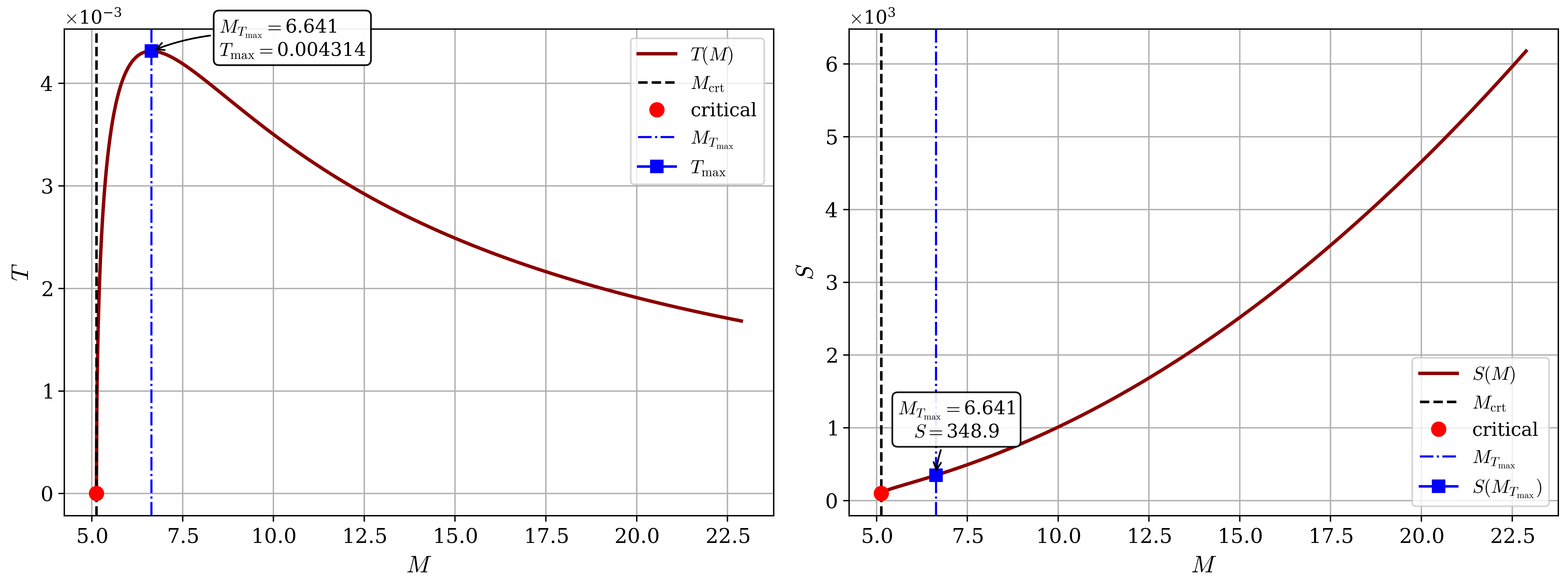} 
	\caption{Temperature and entropy as functions of $M$ for $\sigma \leq \sigma_{\rm crt}$.}
	\label{fig:temp}
\end{figure}

Notice that for the degenerate case $r_{+}^{2}=\tilde{r}^{2}=2c$, we get
$T=0$, that is the extremality condition.  The plots of the temperature and
the entropy in terms of $M$ are given in Fig.~ \ref{fig:temp}
for\footnote{These quantities are defined only over the horizon.}
$\sigma \leq \sigma_{\rm crt}$, that is $M\geq \tilde{M}$.

At large masses the temperature goes like $T\sim 1/M$, which is the typical
Schwarzschild behaviour, but nonlinearities create a peak at
$M_{T_{\rm max}}= 6.641$. Below this mass, $T$ decreases and eventually goes
to zero in the extremal limit. The entropy always increases with the mass and
even in the extremal case is nonzero.

\begin{figure}[!]
	\centering
	\includegraphics[width=0.7\textwidth]{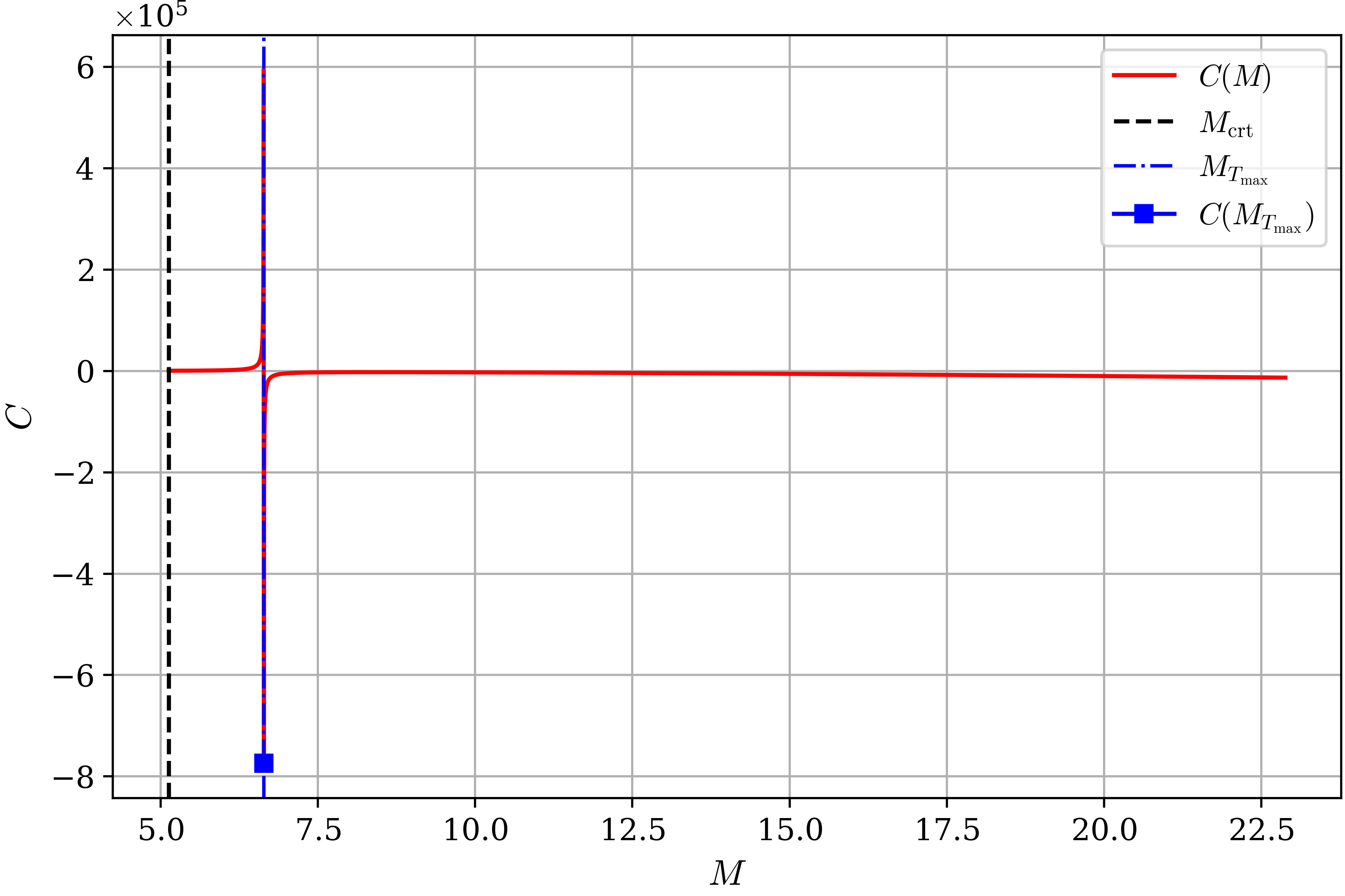} 
	\caption{Specific heat as a function of $M$ for $\sigma \leq \sigma_{\rm crt}$.}
	\label{fig}
\end{figure}

We can also plot the specific heat 

\begin{equation}
	C \equiv T \frac{\partial S}{\partial T} = T \frac{{\partial S}/{\partial M}}{{\partial T}/{\partial M}}
\end{equation}

\noindent
as a function of the mass (see Fig.~\ref{fig}). As shown in
Fig.~\ref{fig:temp}, immediately above the critical mass the temperature
increases, implying $\partial T / \partial M > 0$ and therefore a positive
heat capacity. This behavior is the opposite of that exhibited by the
Schwarzschild solution.  For larger masses the temperature decreases,
therefore $\partial T / \partial M < 0$ and the heat capacity becomes
negative, as in the Schwarzschild solution. At $M_{T_{\rm max}}$ the
temperature has a maximum, so $\partial T / \partial M = 0$ and the specific
heat blows up (phase transition).

\paragraph{Energy conditions and acausality}

We assume the following decomposition of the energy momentum tensor

\begin{equation}\label{energymomentum}
  T^{\mu\nu}
  =
  -\rho {e}_{0}{}^{\mu}{e}_{0}{}^{\nu}
	-\sum_{i} p_{i} {e}^{\;\mu}_{i} {e}^{\;\nu}_{i}\,,
\end{equation}

\noindent
where ${e}_{a}{}^{\mu}$ form an orthonormal basis of vector fields. A
convenient one is given by

\begin{equation}
  e_{0}{}^{\mu}=\bigg(\frac{1}{\sqrt{\lambda}},0,0,0\bigg),\quad
  e_{1}{}^{\mu}=\bigl(0,\sqrt{\lambda},0,0\bigr),\quad
  e_{2}{}^{\mu}=\bigg(0,0,\frac{1}{r},0\bigg),\quad
  e_{3}{}^{\mu}=\bigg(0,0,0,\frac{1}{r\sin\theta}\bigg)\,.
\end{equation}

Defining the orthonormal components of the energy-momentum tensor,
$ T_{ a b}= e_{a}^\mu e_b^\nu T_{\mu\nu}$, using the Einstein's equations and
Eq.~\eqref{matrix}, we can easily read the energy density and the pressure
along the $i$-th direction, that is $ T_{00}=-\rho,\; T_{ i i}=-p_{i}$

\begin{align}
  \rho(r) &= \frac{3 a c}{8\pi G_{N}^{(4)}\,(r^{2}+c)^{5/2}},\\
  \nonumber\\
  p_{1}(r) &= -\frac{3 a c}{8\pi G_{N}^{(4)}\,(r^{2}+c)^{5/2}} = -\rho(r),\\\nonumber\\
  p_{2}(r)=p_{3}(r)
          &= \frac{3 a c\,(3r^{2}-2c)}{16\pi G_{N}^{(4)}\,(r^{2}+c)^{7/2}}\,,
\end{align}

\noindent
where we introduced 

\begin{equation}
  a \equiv 2MG_{N}^{(4)}>0
\end{equation}

\noindent
for simplicity.

In our conventions, the energy conditions \cite{Poisson:2009pwt} are defined
in Table \ref{table}.
	\begin{table}[h]
	\centering
	\begin{tabular}{l|l|l}
		\hline
		\hline
		Name & Statement & Conditions \\
		\hline
		\hline
		Weak (WEC) &
		\(T_{\mu\nu} v^\mu v^\nu \le 0\) &
		\(\rho \ge 0,\ \rho + p_{i} \ge 0\) \\
		\hline
		Null (NEC) &
		\(T_{\mu\nu} k^\mu k^\nu \le 0\) &
		\(\rho + p_{i} \ge 0\) \\
		\hline
		Strong (SEC) &
		\(\bigl(T_{\mu\nu}-\tfrac{1}{2}T\,g_{\mu\nu}\bigr)v^\mu v^\nu \le 0\) &
		\(\rho + p_{i} \ge 0,\ \rho + \sum_{i} p_{i} \ge 0\) \\
		\hline
		Dominant (DEC) &
		\(S^\mu \equiv -T^\mu{}_{\nu}v^\nu\) (future-directed) &
		\(\rho \ge 0,\ \rho \ge |p_{i}|\) \\
		\hline
		\hline
	\end{tabular}
	\caption{Energy conditions in our conventions.}
	\label{table}
\end{table} 

\begin{figure}[h]
	\centering
	\includegraphics[width=0.9\textwidth]{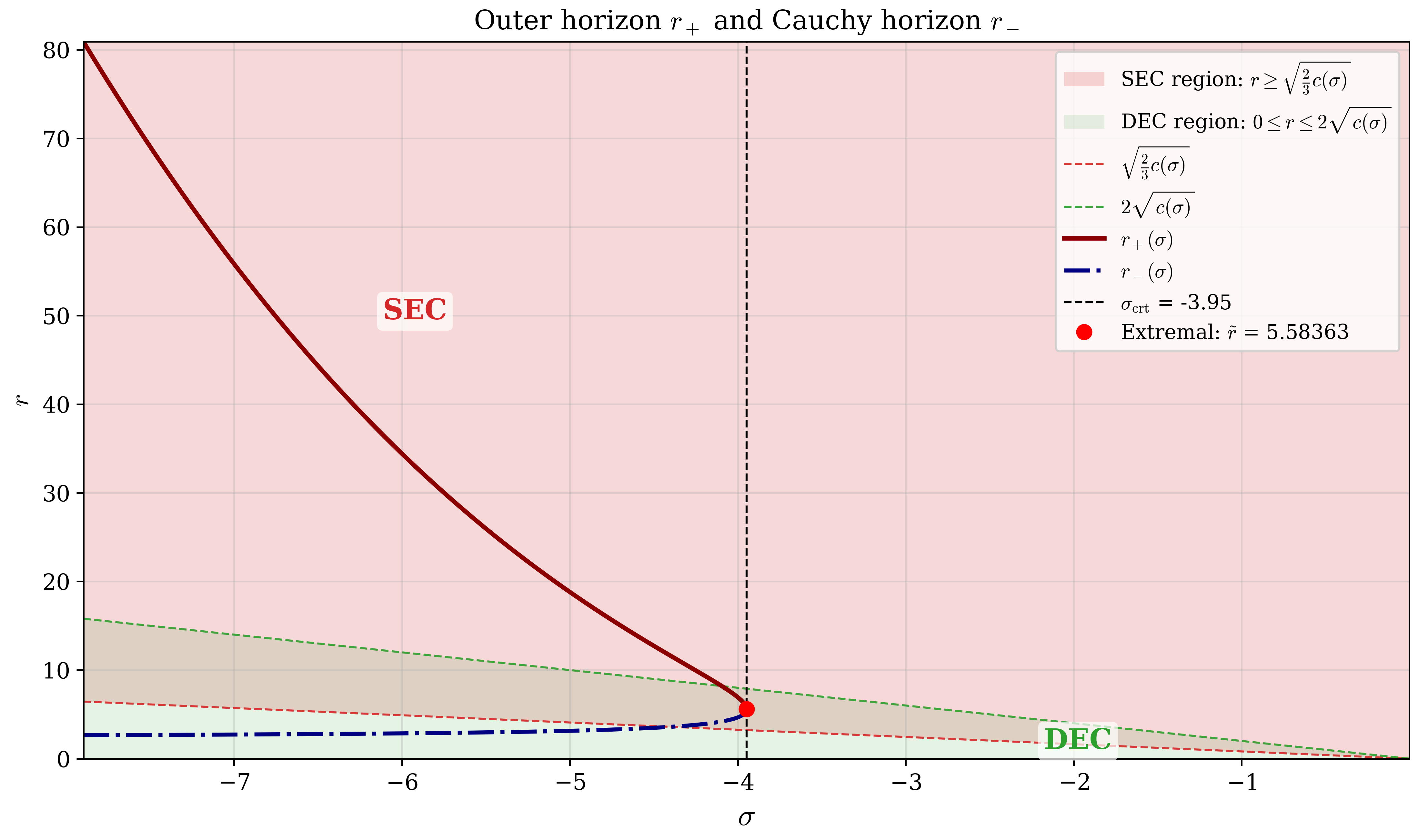} 
	\caption{Outer horizon as a function of $\sigma$.}
	\label{fig3}
\end{figure}

\noindent
The results for the Bardeen solution \cite{Rodrigues:2018bdc} are
collected in Table~\ref{table2}. The null energy condition (NEC) and
the weak energy condition (WEC) are satisfied everywhere, while the
strong energy condition (SEC) holds only for radii
\(r>\sqrt{\tfrac{2c}{3}}\). This behaviour is expected from the
Penrose--Hawking singularity theorems
\cite{Penrose:1964wq,Hawking:1970zqf}:\footnote{For an interesting
  review with a historical perspective and many related references,
  see Ref.~\cite{Senovilla:2014gza}.} since the Bardeen solution is
regular, the SEC must be violated in the region where a singularity
would otherwise occur, i.e. near $r=0$. The inner and outer horizons,
together with the radii delimiting the regions where the energy
conditions hold are shown in Fig.~\ref{fig3} as functions of
\(\sigma\).

	\begin{table}[h]
		\centering
		\begin{tabular}{cc}
			\hline
			\hline
			Energy condition & Satisfied in\\
			\hline
			\hline
			NEC  & Everywhere \\
			WEC  & Everywhere\\
			SEC  & $ r \geq \sqrt{\frac{2}{3}c}$ \\
			DEC  & $0\leq r \leq  2\sqrt{c}$ \\
			\hline
			\hline
		\end{tabular}
		\caption{Energy conditions for the Bardeen solution.}
		\label{table2}
	\end{table}

In Ref.~\cite{Russo:2024xnh} it was shown that the SEC can be violated only
for acausal NED theories.  Several properties of black-hole solutions in
Einstein--NED models can be characterized by specific functions of the radial
coordinate $r$.  Of particular interest here is the relation between the
energy conditions, in this case the SEC, and causality.  In
Ref.~\cite{Hale:2025ezt} a function $\mathcal{E}(r)$, interpreted as the
electromagnetic energy outside a sphere of radius $r$, was introduced. Its
definition coincides with the function $f(r;\alpha)$ given in
Eq.~\eqref{lambda}.  The SEC, which is a necessary but not sufficient
condition for causality, implies that $\mathcal{E}(r)$ must be strictly convex
in $r$.  This can be seen with the following argument: from Einstein's
equations we know that

\begin{equation}
	R_{tt}= -\frac{\lambda G_{N}^{(4)} }{r}\mathcal{E}''(r)\,,
\end{equation}

\noindent
and that the SEC implies $R_{tt}\leq 0 $ (see Table \ref{table}). It follows
that $\mathcal{E}''(r)\geq 0$.  In Ref.~\cite{Russo:2026vnj} another relevant
function has been introduced: the effective charge $Q_{\mathrm{eff}}^{2}(r)$
defined by

\begin{equation}
2\pi\,Q_{\mathrm{eff}}^{2}(r) \equiv r\,\mathcal{E}(r).
\end{equation}

\noindent
This function captures the nonlinear modification of the Reissner\textendash
Nordstr\"om charge that appears in the metric function. It has been shown that
$Q_{\mathrm{eff}}^{2}(r)$ must be convex\footnote{In \cite{Russo:2026vnj} they
  proved it must be concave, but in our conventions the trace of the
  energy-momentum tensor is positive, so $Q_{\mathrm{eff}}^{2}(r)$ must be
  convex.} in $r$ in order to ensure causality.

In our case the function $\mathcal{E}(r)$ is given by Eq.~\eqref{f}:

\begin{equation}
  \mathcal{E}(r)
  =
  M- \frac{Mr^{3}}{[r^{2}+\sqrt{\alpha}\,\sigma^{2}]^{3/2}}.
\end{equation}
      
Computing the second derivative of $\mathcal{E}$ and $Q^{2}_{\rm eff}$ we
easily see that

\begin{align}
	\text{Causality} \;\qquad&\Rightarrow\; \qquad\mathcal{E}''(r) \ge 0
	&&\Rightarrow\; \qquad r \ge \sqrt{\tfrac{2}{3}c}, \label{eq:E}\\
	\text{Causality} \;\qquad&\Rightarrow\; \qquad\bigl(Q_{\mathrm{eff}}^{2}(r)\bigr)'' \ge 0
	&&\Rightarrow\; \qquad r \ge 2\sqrt{c}. \label{eq:Q}
\end{align}

Therefore, the region where the SEC is not satisfied is
$0 \leq r \leq \sqrt{\frac{2}{3} c}$ and the theory is acausal, as expected
for our regular Bardeen solution.

To conclude, we seek for a relation between the function $\mathcal{E}(r)$
introduced in Ref.~\cite{Hale:2025ezt} and the Komar integrals in
Eqs.~\eqref{tildePF} and \eqref{alphaP}, thereby providing a physical
interpretation of the latter.  The integral \eqref{eep} is equal to
$ r^{2} \lambda'/(4G^{(4)}_{N})$, and we can write it in terms of
$\mathcal{E}(r)$. On the other hand, the sum of the three integrals must give
$M/2$, so we can easily deduce that

\begin{equation}
	\frac{1}{16\pi G_{N}^{(4)}}\int
	\bigg\{	-	\tfrac{1}{2}\tilde{P}_{k}  F
	+\frac{\gamma}{2}\alpha P_{k}^{H}
	\bigg\} = \frac{\mathcal{E}(r)-r\mathcal{E}'(r)}{2}= \tfrac{1}{2} \bigg(\mathcal{E}(r)+4\pi r^{3}\rho \bigg)\,,
\end{equation}

\noindent
which gives the interpretation of the last two terms of the Komar integral in
terms of the energy functions defined above.

The conditions for the existence of an event horizon and some thermodynamical
properties of the black holes in a class of causal theories of Einstein--NLED
has also been studied in Ref.~\cite{Babaei-Aghbolagh:2025tim}.

\section{Discussion}\label{sectionconclusion}

In this article we have derived a generalized Komar charge for
Einstein gravity coupled to generic theories of non-linear
electrodynamics in 4 dimensions and we have used it to derive the
Smarr formula and the first law of black-hole thermodynamics. We have
also used it to study the mechanism underlying the absence of
singularities in black-hole and soliton solutions showing how the
entire mass of these objects is sourced by the non-linear
electromagnetic field. We have also studied how the existence of
regular black holes is associated to violations of energy conditions
(as the singularity theorems \cite{Penrose:1964wq,Hawking:1970zqf}
indicate) and to the acausal nature of the theory of NLED under
consideration \cite{Russo:2024xnh,Russo:2026vnj}.

While the astrophysical relevance of the regular solitons and black
holes of the Einstein--NLED theories is unclear, they provide a very
interesting theoretical playing ground on which ideas, general results
can be tested and interesting questions can be asked. A very
interesting question we would like to ask is what is the mechanism
that can produce the transition between a regular black hole and a
soliton in some of these theories. Why is a horizon needed in some
regions of the parameter space and not in others, if there are no
singularities to be hidden by the cosmic censor? Penrose's singularity
theorem leads to the inevitable existence of black holes only when we
combine it with the cosmic censorship hypothesis. Can gravitational
collapse in these theories lead to horizonless solitons or to regular
black holes depending on the initial data? More work is necessary to
address these important questions.

\section*{Acknowledgments}

The work of GB and TO has been supported in part by the MCI, AEI,
FEDER (UE) grant PID2024-155685NB-C21 ``Gravity, Supergravity and
Superstrings'' (GRASS) and ``IFT Centro de Excelencia Severo Ochoa''
CEX2020-001007-S. The work of GB has been supported by the fellowship
CEX2020-001007-S-20-5. TO wishes to thank M.M.~Fern\'andez for her
permanent support.

\appendix
\section{Appendix }\label{Appendix}

\subsection{The magnetic Reissner--Nordstr\"om black hole}

The formalism developed in the previous sections applies to all nonlinear
electrodynamics theories, including the standard linear Maxwell theory, in
which the coupling constant vanishes and the Lagrangian reduces to the usual
Maxwell form:

\begin{equation}
	\mathscr{L} =X\,.
\end{equation}
      
Here we want to consider the magnetic configuration, hence using Eq.~\eqref{X}
we have

\begin{align}
  \mathscr{L}&\doteq  - \frac{\sigma^{4}}{r^{4}}\,,
  \\\nonumber\\
  \mathscr{L}_{\alpha}  &= 0\,,
  \\\nonumber\\
  \mathscr{L}_{X} & = 1\,.
\end{align}

We can find $f$ integrating Eq.~\eqref{lf'}

\begin{equation}
  f(r;\alpha)
  \doteq
  \frac{\sigma^{4}}{4G^{(4)}_{N}r} +c.
\end{equation} 

Setting the integration constant $c= 0 $ ensures
$ f(r;\alpha)\stackrel{r\rightarrow\infty}{\rightarrow}{0}$.

The second integral in Eq.~\eqref{magneticsmarr}  gives

\begin{equation}
	\int_{r_{+}}^{\infty}\frac{1}{ r^{2}}\mathscr{L}_{X} \dd r \doteq
	\frac{1}{r_{+}}\,.
\end{equation}

Combining these results it is easy to check that Eq.~\eqref{magneticsmarr} is
satisfied, and therefore we verified the Smarr formula \eqref{smarrrrrr}.

We can also compute each term of the Komar integral. The first one is

\begin{equation}\label{RNkomar}
\frac{1}{16\pi G_{N}^{(4)}}
\int_{S^{2}_{r}}\big\{-\star (e^{a} \wedge e^{b})P_{k ab}\big\}
=
\frac{M}{2}-\frac{\sigma^{4}}{4G^{(4)}_{N}r}\,.
\end{equation}

Using Eq.~\eqref{tildeP}, we get the dual-momentum map as a function of $r$

\begin{equation}
	\tilde{P}_{k}(r) \doteq -\frac{4G^{(4)}_{N}\mathcal{P}}{r}\,,
\end{equation}

\noindent
where we chose the integration constant to get
$\tilde{P}_{k}(r) \stackrel{r\rightarrow\infty}{\rightarrow}0$. With this
choice, the second term in Eq.~\eqref{komarbardeen} gives

\begin{equation}
  \label{RNPtilde}
  \frac{1}{16\pi G_{N}^{(4)}}
  \int_{S^{2}_{r}}\bigg\{ - \tfrac{1}{2}\tilde{P}_{k}(r) F\bigg\}
  =
  \frac{\sigma^{4}}{4G^{(4)}_{N}r}\,.
\end{equation}

Combining Eqs.~\eqref{RNkomar} and \eqref{RNPtilde} we get that the Komar
integral is $M/2$, independently of the radius, as expected from Stokes
theorem.



\begin{thebibliography}{99}
	
\bibitem{Dymnikova:1992ux}
I.~Dymnikova,
``Vacuum nonsingular black hole,''
Gen. Rel. Grav. \textbf{24} (1992), 235-242
\doi{10.1007/BF00760226}

\bibitem{Ayon-Beato:1998hmi}
E.~Ay\'on-Beato and A.~Garc\'{\i}a,
``Regular black hole in general relativity coupled to nonlinear electrodynamics,''
Phys. Rev. Lett. \textbf{80} (1998), 5056-5059
\doi{10.1103/PhysRevLett.80.5056}
[\grqc{9911046} [gr-qc]].

\bibitem{Ayon-Beato:1999kuh}
E.~Ay\'on-Beato and A.~Garc\'{\i}a,
``New regular black hole solution from nonlinear electrodynamics,''
Phys. Lett. B \textbf{464} (1999), 25
\doi{10.1016/S0370-2693(99)01038-2}
[\hepth{9911174} [hep-th]].

\bibitem{Ayon-Beato:1999qin}
E.~Ay\'on-Beato and A.~Garc\'{\i}a,
``Nonsingular charged black hole solution for nonlinear source,''
Gen. Rel. Grav. \textbf{31} (1999), 629-633
\doi{10.1023/A:1026640911319}
[\grqc{9911084} [gr-qc]].

\bibitem{Ayon-Beato:2000mjt}
E.~Ay\'on-Beato and A.~Garc\'{\i}a,
``The Bardeen model as a nonlinear magnetic monopole,''
Phys. Lett. B \textbf{493} (2000), 149-152
\doi{10.1016/S0370-2693(00)01125-4}
[\grqc{0009077} [gr-qc]].

\bibitem{Ayon-Beato:2004ywd}
E.~Ayon-Beato and A.~Garcia,
``Four parametric regular black hole solution,''
Gen. Rel. Grav. \textbf{37} (2005), 635
\doi{10.1007/s10714-005-0050-y}
[\hepth{0403229} [hep-th]].

\bibitem{Bronnikov:2000vy}
K.~A.~Bronnikov,
``Regular magnetic black holes and monopoles from nonlinear electrodynamics,''
Phys. Rev. D \textbf{63} (2001), 044005
\doi{10.1103/PhysRevD.63.044005}
[\grqc{0006014} [gr-qc]].

\bibitem{Dymnikova:2015hka}
I.~Dymnikova and E.~Galaktionov,
``Regular rotating electrically charged black holes and solitons in non-linear electrodynamics minimally coupled to gravity,''
Class. Quant. Grav. \textbf{32} (2015) no.16, 165015
\doi{10.1088/0264-9381/32/16/165015}
[\arxiv{1510.01353} [gr-qc]].

\bibitem{Bronnikov:2022ofk}
K.~A.~Bronnikov,
``Regular black holes sourced by nonlinear electrodynamics,''
[\arxiv{2211.00743} [gr-qc]].

\bibitem{Bronnikov:2017sgg}
K.~A.~Bronnikov,
``Nonlinear electrodynamics, regular black holes and wormholes,''
Int. J. Mod. Phys. D \textbf{27} (2018) no.06, 1841005
\doi{10.1142/S0218271818410055}
[\arxiv{1711.00087} [gr-qc]].

\bibitem{Dymnikova:2004zc}
I.~Dymnikova,
``Regular electrically charged structures in nonlinear electrodynamics coupled to general relativity,''
Class. Quant. Grav. \textbf{21} (2004), 4417-4429
\doi{10.1088/0264-9381/21/18/009}
[\grqc{0407072} [gr-qc]].

\bibitem{Smarr:1972kt}
L.~Smarr,
``Mass formula for Kerr black holes,''
Phys. Rev. Lett. \textbf{30} (1973), 71-73
[erratum: Phys. Rev. Lett. \textbf{30} (1973), 521-521]
\doi{10.1103/PhysRevLett.30.71}
  
\bibitem{Rasheed:1997ns}
D.~A.~Rasheed,
``Nonlinear electrodynamics: Zeroth and first laws of black hole mechanics,''
[\hepth{9702087} [hep-th]].

\bibitem{Breton:2004qa}
N.~Breton,
``Smarr's formula for black holes with non-linear electrodynamics,''
Gen. Rel. Grav. \textbf{37} (2005), 643-650
\doi{10.1007/s10714-005-0051-x}
[\grqc{0405116} [gr-qc]].

\bibitem{Gonzalez:2009nn}
H.~A.~Gonz\'alez, M.~Hassaine and C.~Mart\'{\i}nez,
``Thermodynamics of charged black holes with a nonlinear electrodynamics source,''
Phys. Rev. D \textbf{80} (2009), 104008
\doi{10.1103/PhysRevD.80.104008}
[\arxiv{0909.1365} [hep-th]].

\bibitem{Yi-Huan:2010jnv}
W.~Yi-Huan,
``Energy and first law of thermodynamics for Born-Infeld-anti-de-Sitter black hole,''
Chin. Phys. B \textbf{19} (2010), 090404
\doi{10.1088/1674-1056/19/9/090404}

\bibitem{Gunasekaran:2012dq}
S.~Gunasekaran, R.~B.~Mann and D.~Kubiznak,
``Extended phase space thermodynamics for charged and
rotating black holes and Born-Infeld vacuum polarization,''
JHEP \textbf{11} (2012), 110
\doi{10.1007/JHEP11(2012)110}
[\arxiv{1208.6251} [hep-th]].

\bibitem{Diaz-Alonso:2012lkh}
J.~D\'{\i}az-Alonso and D.~Rubiera-Garc\'{\i}a,
``Thermodynamic analysis of black hole solutions
in gravitating nonlinear electrodynamics,''
Gen. Rel. Grav. \textbf{45} (2013), 1901-1950
\doi{10.1007/s10714-013-1567-0}
[\arxiv{1204.2506} [gr-qc]].

\bibitem{Zhang:2016ilt}
Y.~Zhang and S.~Gao,
``First law and Smarr formula of black hole mechanics in
nonlinear gauge theories,''
Class. Quant. Grav. \textbf{35} (2018) no.14, 145007
\doi{10.1088/1361-6382/aac9d4}
[\arxiv{1610.01237} [gr-qc]].

\bibitem{Fan:2016hvf}
Z.~Y.~Fan and X.~Wang,
``Construction of Regular Black Holes in General Relativity,''
Phys. Rev. D \textbf{94} (2016) no.12, 124027
\doi{10.1103/PhysRevD.94.124027}
[\arxiv{1610.02636} [gr-qc]].

\bibitem{Hu:2018njr}
S.~Q.~Hu, X.~M.~Kuang and Y.~C.~Ong,
``A Note on Smarr Relation and Coupling Constants,''
Gen. Rel. Grav. \textbf{51} (2019) no.5, 55
\doi{10.1007/s10714-019-2540-3}
[\arxiv{1810.06073} [gr-qc]].

\bibitem{Kastor:2009wy}
D.~Kastor, S.~Ray and J.~Traschen,
``Enthalpy and the Mechanics of AdS Black Holes,''
Class. Quant. Grav. \textbf{26} (2009), 195011
\doi{10.1088/0264-9381/26/19/195011}
[\arxiv{0904.2765} [hep-th]].

\bibitem{Kubiznak:2016qmn}
D.~Kubiznak, R.~B.~Mann and M.~Teo,
``Black hole chemistry: thermodynamics with Lambda,''
Class. Quant. Grav. \textbf{34} (2017) no.6, 063001
\doi{10.1088/1361-6382/aa5c69}
[\arxiv{1608.06147} [hep-th]].

\bibitem{Gulin:2017ycu}
L.~Gulin and I.~Smoli{\'c},
``Generalizations of the Smarr formula for black holes
with nonlinear electromagnetic fields,''
Class. Quant. Grav. \textbf{35} (2018) no.2, 025015
\doi{10.1088/1361-6382/aa9dfd}
[\arxiv{1710.04660} [gr-qc]].

\bibitem{Townsend:1997ku}
P.~K.~Townsend,
``Black holes: Lecture notes,''
[\grqc{9707012} [gr-qc]].

\bibitem{Bokulic:2021dtz}
A.~Bokuli{\'c}, T.~Juri{\'c} and I.~Smoli{\'c},
``Black hole thermodynamics in the presence of nonlinear electromagnetic fields,''
Phys. Rev. D \textbf{103} (2021) no.12, 124059
\doi{10.1103/PhysRevD.103.124059}
[\arxiv{2102.06213} [gr-qc]].

\bibitem{Balart:2017dzt}
L.~Balart and S.~Fernando,
``A Smarr formula for charged black holes in nonlinear electrodynamics,''
Mod. Phys. Lett. A \textbf{32} (2017) no.39, 1750219
\doi{10.1142/S0217732317502194}
[\arxiv{1710.07751} [gr-qc]].

\bibitem{Bardeen:1973gs}
J.~M.~Bardeen, B.~Carter and S.~W.~Hawking,
``The Four laws of black hole mechanics,''
Commun. Math. Phys. \textbf{31} (1973), 161-170
\doi{10.1007/BF01645742}


\bibitem{Komar:1958wp}
A.~Komar,
``Covariant conservation laws in general relativity,''
Phys. Rev. \textbf{113} (1959), 934-936
\doi{10.1103/PhysRev.113.934}


\bibitem{Barbagallo:2025qdy}
G.~Barbagallo, J.~L.~V.~Cerdeira, C.~G\'omez-Fayr\'en,
P.~Meessen and T.~Ort\'{\i}n,
``Higher-form symmetries in supergravity, scalar charges and
black-hole thermodynamics,''
[\arxiv{2512.22565} [hep-th]]. To be published in JHEP.

\bibitem{Carter:1973rla}
B.~Carter,
``Black holes equilibrium states,''
Contribution to: Les Houches Summer School of Theoretical Physics, 57-214.

\bibitem{Magnon:1985sc}
A.~Magnon,
``On Komar integrals in asymptotically anti-de Sitter space-times,''
J. Math. Phys. \textbf{26} (1985), 3112-3117
\doi{10.1063/1.526690}

\bibitem{Bazanski:1990qd}
S.~L.~Bazanski and P.~Zyla,
``A Gauss type law for gravity with a cosmological constant,''
Gen. Rel. Grav. \textbf{22} (1990), 379-387

\bibitem{Kastor:2008xb}
D.~Kastor,
``Komar Integrals in Higher (and Lower) Derivative Gravity,''
Class. Quant. Grav. \textbf{25} (2008), 175007
\doi{10.1088/0264-9381/25/17/175007}
[\arxiv{0804.1832} [hep-th]].

\bibitem{Kastor:2010gq}
D.~Kastor, S.~Ray and J.~Traschen,
``Smarr Formula and an Extended First Law for Lovelock Gravity,''
Class. Quant. Grav. \textbf{27} (2010), 235014
\doi{10.1088/0264-9381/27/23/235014}
[\arxiv{1005.5053} [hep-th]].

\bibitem{Ortin:2021ade}
T.~Ort\'{\i}n,
``Komar integrals for theories of higher order
in the Riemann curvature and black-hole chemistry,''
JHEP \textbf{08} (2021), 023
\doi{10.1007/JHEP08(2021)023}
[\arxiv{2104.10717} [gr-qc]].

\bibitem{Meessen:2022hcg}
P.~Meessen, D.~Mitsios and T.~Ort\'{\i}n,
``Black hole chemistry, the cosmological constant and the embedding tensor,''
JHEP \textbf{12} (2022), 155
\doi{10.1007/JHEP12(2022)155}
[\arxiv{2203.13588} [hep-th]].

\bibitem{Liberati:2015xcp}
S.~Liberati and C.~Pacilio,
``Smarr Formula for Lovelock Black Holes: a Lagrangian approach,''
Phys. Rev. D \textbf{93} (2016) no.8, 084044
\doi{10.1103/PhysRevD.93.084044}
[\arxiv{1511.05446} [gr-qc]].

\bibitem{Lee:1990nz}
J.~Lee and R.~M.~Wald,
``Local symmetries and constraints,''
J. Math. Phys. \textbf{31} (1990), 725-743
\doi{10.1063/1.528801}

\bibitem{Wald:1993nt}
R.~M.~Wald,
``Black hole entropy is the Noether charge,''
Phys. Rev. D \textbf{48} (1993) no.8, R3427-R3431
\doi{10.1103/PhysRevD.48.R3427}
[\grqc{9307038} [gr-qc]].

\bibitem{Elgood:2020svt}
Z.~Elgood, P.~Meessen and T.~Ort\'{\i}n,
``The first law of black hole mechanics in
the Einstein-Maxwell theory revisited,''
JHEP \textbf{09} (2020), 026
\doi{10.1007/JHEP09(2020)026}
[\arxiv{2006.02792} [hep-th]].

\bibitem{Elgood:2020mdx}
Z.~Elgood, D.~Mitsios, T.~Ort\'{\i}n and D.~Peren\'{\i}guez,
``The first law of heterotic stringy black hole mechanics
at zeroth order in {\ensuremath{\alpha}}',''
JHEP \textbf{07} (2021), 007
\doi{10.1007/JHEP07(2021)007}
[\arxiv{2012.13323} [hep-th]].

\bibitem{Elgood:2020nls}
Z.~Elgood, T.~Ort\'{\i}n and D.~Peren\'{\i}guez,
``The first law and Wald entropy formula of heterotic
stringy black holes at first order in $\alpha'$,''
JHEP \textbf{05} (2021), 110
\doi{10.1007/JHEP05(2021)110}
[\arxiv{2012.14892} [hep-th]].

\bibitem{Mitsios:2021zrn}
D.~Mitsios, T.~Ort\'{\i}n and D.~Pere\~niguez,
``Komar integral and Smarr formula for axion-dilaton black holes versus S duality,''
JHEP \textbf{08} (2021), 019
\doi{10.1007/JHEP08(2021)019}
[\arxiv{2106.07495} [hep-th]].

\bibitem{Ortin:2022uxa}
T.~Ort\'{\i}n and D.~Pere\~niguez,
``Magnetic charges and Wald entropy,''
JHEP \textbf{11} (2022), 081
\doi{10.1007/JHEP11(2022)081}
[\arxiv{2207.12008} [hep-th]].

\bibitem{Misner:1957mt}
C.~W.~Misner and J.~A.~Wheeler,
``Classical physics as geometry: Gravitation,
electromagnetism, unquantized charge, and mass as properties of curved empty space,''
Annals Phys. \textbf{2} (1957), 525-603
\doi{10.1016/0003-4916(57)90049-0}

\bibitem{Ballesteros:2023muf}
R.~Ballesteros and T.~Ort\'{\i}n,
``Hairy black holes, scalar charges and extended thermodynamics,''
Class. Quant. Grav. \textbf{41} (2024) no.5, 055007
\doi{10.1088/1361-6382/ad210a}
[\arxiv{2308.04994} [gr-qc]].

\bibitem{Ballesteros:2024prz}
R.~Ballesteros and T.~Ort\'{\i}n,
``Generalized Komar charges and Smarr formulas
for black holes and boson stars,''
SciPost Phys. Core \textbf{8} (2025) no.038, 038
\doi{10.21468/SciPostPhysCore.8.2.038}
[\arxiv{2409.08268} [gr-qc]].

\bibitem{Wald:1990mme}
R.~M.~Wald,
``On identically closed forms locally constructed from a field,''
J. Math. Phys. \textbf{31} (1990) no.10, 2378
\doi{10.1063/1.528839}

\bibitem{Ballesteros:2025wvs}
R.~Ballesteros, M.~C\'ardenas and E.~Lescano,
``Nonclosed scalar charge in four-dimensional
Einstein-scalar-Gauss-Bonnet black hole thermodynamics,''
Phys. Rev. D \textbf{113} (2026) no.8, 084039
\doi{10.1103/1hz8-zn5h}
[\arxiv{2510.08686} [hep-th]].

\bibitem{Cremmer:1980gs}
E.~Cremmer,
``Supergravities in 5 Dimensions,''
LPTENS-80-17.

\bibitem{Born:1934gh}
M.~Born and L.~Infeld,
``Foundations of the new field theory,''
Proc. Roy. Soc. Lond. A \textbf{144} (1934) no.852, 425-451
\doi{10.1098/rspa.1934.0059}

\bibitem{Gibbons:2013tqa}
G.~W.~Gibbons and N.~P.~Warner,
``Global structure of five-dimensional fuzzballs,''
Class. Quant. Grav. \textbf{31} (2014), 025016
\doi{10.1088/0264-9381/31/2/025016}
[\arxiv{1305.0957} [hep-th]].

\bibitem{Cano:2018aod}
P.~A.~Cano, S.~Chimento, T.~Ort\'{\i}n and A.~Ruip\'erez,
``Regular Stringy Black Holes?,''
Phys. Rev. D \textbf{99} (2019) no.4, 046014
\doi{10.1103/PhysRevD.99.046014}
[\arxiv{1806.08377} [hep-th]].

\bibitem{Yazadjiev:2024rql}
S.~Yazadjiev and D.~Doneva,
``Black hole no-hair theorem for self-gravitating time-dependent
spherically symmetric multiple scalar fields,''
Eur. Phys. J. C \textbf{84} (2024) no.5, 492
\doi{10.1140/epjc/s10052-024-12822-6}
[\arxiv{2401.13288} [gr-qc]].

\bibitem{Smolic:2015txa}
I.~Smoli\'c,
``Symmetry inheritance of scalar fields,''
Class. Quant. Grav. \textbf{32} (2015) no.14, 145010
\doi{10.1088/0264-9381/32/14/145010}
[\arxiv{1501.04967} [gr-qc]].

\bibitem{Bokulic:2025brf}
A.~Bokuli{\'c}, T.~Juri{\'c} and I.~Smoli{\'c},
``Conundrum of regular black holes with nonlinear electromagnetic fields,''
Phys. Rev. D \textbf{113} (2026) no.2, 024044
\doi{10.1103/z7gd-96ms}
[\arxiv{2510.23711} [gr-qc]].

\bibitem{Ortin:2015hya}
T.~Ort\'{\i}n,
``Gravity and Strings'', 2nd edition, 
Cambridge University Press, 2015.

\bibitem{Sorokin:2021tge}
D.~P.~Sorokin,
``Introductory Notes on Non-linear Electrodynamics and its Applications,''
Fortsch. Phys. \textbf{70} (2022) no.7-8, 2200092
\doi{10.1002/prop.202200092}
[\arxiv{2112.12118} [hep-th]].

\bibitem{Cerdeira:2025elp}
J.~L.~V.~Cerdeira and T.~Ort\'{\i}n,
``On-shell Lagrangians as total derivatives and the generalized Komar charge,''
JHEP \textbf{09} (2025), 068
\doi{10.1007/JHEP09(2025)068}
[\arxiv{2506.14024} [hep-th]].

\bibitem{Golshani:2024fry}
M.~Golshani, M.~M.~Sheikh-Jabbari,
V.~Taghiloo and M.~H.~Vahidinia,
``Reloading Black Hole Thermodynamics with Noether Charges,''
[\arxiv{2407.15994} [hep-th]].

\bibitem{Adami:2024gdx}
H.~Adami, M.~Golshani, M.~M.~Sheikh-Jabbari,
V.~Taghiloo and M.~H.~Vahidinia,
``Covariant phase space formalism for fluctuating boundaries,''
JHEP \textbf{09} (2024), 157
\doi{10.1007/JHEP09(2024)157}
[\arxiv{2407.03259} [hep-th]].

\bibitem{Ortin:2024mmg}
T.~Ort\'{\i}n and M.~Zatti,
``A note on the Noether-Wald and generalized Komar charges,''
SciPost Phys. Core \textbf{8} (2025), 094
\doi{10.21468/SciPostPhysCore.8.4.094}
[\arxiv{2411.10420} [gr-qc]].

\bibitem{Zatti:2024vqv}
M.~Zatti,
``Exploring String Theory Solutions: Black hole thermodynamics with
$\alpha'$ corrections and type II compactifications'',
PhD thesis, U. Aut\'onoma de Madrid, 2024.
[\arxiv{2505.01191} [hep-th]].

\bibitem{Ma:2014qma}
M.~S.~Ma and R.~Zhao,
``Corrected form of the first law of thermodynamics for regular black holes,''
Class. Quant. Grav. \textbf{31} (2014), 245014
\doi{10.1088/0264-9381/31/24/245014}
[\arxiv{1411.0833} [gr-qc]].

\bibitem{Ma:2025dee}
M.~S.~Ma, H.~F.~Li and J.~H.~Shi,
``Regular black holes and reductions of thermodynamic phase spaces,''
Sci. China Phys. Mech. Astron. \textbf{69} (2026) no.1, 210411
\doi{10.1007/s11433-025-2753-6}
[\arxiv{2507.09551} [gr-qc]].

\bibitem{Toshmatov:2018cks}
B.~Toshmatov, Z.~Stuchl{\'\i}k and B.~Ahmedov,
``Comment on {\textquotedblleft}Construction of regular
black holes in general relativity{\textquotedblright},''
Phys. Rev. D \textbf{98} (2018) no.2, 028501
\doi{10.1103/PhysRevD.98.028501}
[\arxiv{1807.09502} [gr-qc]].

\bibitem{Bandos:2020jsw}
I.~Bandos, K.~Lechner, D.~Sorokin and P.~K.~Townsend,
``A non-linear duality-invariant conformal extension of Maxwell's equations,''
Phys. Rev. D \textbf{102} (2020), 121703
\doi{10.1103/PhysRevD.102.121703}
[\arxiv{2007.09092} [hep-th]].

\bibitem{Kosyakov:2020wxv}
B.~P.~Kosyakov,
``Nonlinear electrodynamics with the maximum allowable symmetries,''
Phys. Lett. B \textbf{810} (2020), 135840
\doi{10.1016/j.physletb.2020.135840}
[\arxiv{2007.13878} [hep-th]].




\bibitem{Man:2013hpa}
J.~Man and H.~Cheng,
``The calculation of the thermodynamic quantities of the Bardeen black hole,''
Gen. Rel. Grav. \textbf{46} (2014), 1660
\doi{10.1007/s10714-013-1660-4}
[\arxiv{1304.5686} [hep-th]].

\bibitem{Tzikas:2018cvs}
A.~G.~Tzikas,
``Bardeen black hole chemistry,''
Phys. Lett. B \textbf{788} (2019), 219-224
\doi{10.1016/j.physletb.2018.11.036}
[\arxiv{1811.01104} [gr-qc]].

\bibitem{Poisson:2009pwt}
E.~Poisson,
``A Relativist's Toolkit: The Mathematics of Black-Hole Mechanics,''
Cambridge University Press, 2009,
\doi{10.1017/CBO9780511606601}

\bibitem{Rodrigues:2018bdc}
M.~E.~Rodrigues and M.~V.~de Sousa Silva,
``Bardeen Regular Black Hole With an Electric Source,''
JCAP \textbf{06} (2018), 025
\doi{10.1088/1475-7516/2018/06/025}
[\arxiv{1802.05095} [gr-qc]].

\bibitem{Penrose:1964wq}
R.~Penrose,
``Gravitational collapse and space-time singularities,''
Phys. Rev. Lett. \textbf{14} (1965), 57-59
\doi{10.1103/PhysRevLett.14.57}

\bibitem{Hawking:1970zqf}
S.~W.~Hawking and R.~Penrose,
``The Singularities of gravitational collapse and cosmology,''
Proc. Roy. Soc. Lond. A \textbf{314} (1970), 529-548
\doi{10.1098/rspa.1970.0021}

\bibitem{Senovilla:2014gza}
J.~M.~M.~Senovilla and D.~Garfinkle,
``The 1965 Penrose singularity theorem,''
Class. Quant. Grav. \textbf{32} (2015) no.12, 124008
\doi{10.1088/0264-9381/32/12/124008}
[\arxiv{1410.5226} [gr-qc]].

\bibitem{Russo:2024xnh}
J.~G.~Russo and P.~K.~Townsend,
``Causality and energy conditions in nonlinear electrodynamics,''
JHEP \textbf{06} (2024), 191
\doi{10.1007/JHEP06(2024)191}
[\arxiv{2404.09994} [hep-th]].

\bibitem{Hale:2025ezt}
T.~Hale, R.~A.~Hennigar and D.~Kubiznak,
``Excising Cauchy Horizons with Nonlinear Electrodynamics,''
Phys. Rev. D \textbf{113} (2026) no.6, L061502
\doi{10.1103/x8c8-j85k}
[\arxiv{2506.20802} [gr-qc]].


\bibitem{Russo:2026vnj}
J.~G.~Russo and P.~K.~Townsend,
``Black holes and causal nonlinear electrodynamics,''
[\arxiv{2601.07789} [hep-th]].

\bibitem{Babaei-Aghbolagh:2025tim}
H.~Babaei-Aghbolagh, K.~Babaei Velni, S.~He and F.~Isapour,
``Unified causal framework for a nonlinear electrodynamics black hole
from the Courant-Hilbert approach: Thermodynamics and singularity,''
Phys. Rev. D \textbf{113} (2026) no.8, 084041
\doi{10.1103/zfc2-52md}
[\arxiv{2511.17407} [hep-th]].











\end{thebibliography}
\end{document}